%% file: main.tex
\documentclass[manuscript]{acmart}
\usepackage{color}

\usepackage{graphicx}
\usepackage{subfigure}

\usepackage{booktabs}
\usepackage{multirow}

\usepackage[toc,page]{appendix} 
\usepackage{tabularx}
\usepackage{url}
\usepackage{enumitem}
\setlist[itemize]{leftmargin=*}

\usepackage[ruled]{algorithm2e}

\AtBeginDocument{%
  \providecommand\BibTeX{{%
    \normalfont B\kern-0.5em{\scshape i\kern-0.25em b}\kern-0.8em\TeX}}}

\begin{document}


\title{Who Should I Trust: AI or Myself? Leveraging Human and AI Correctness Likelihood to Promote Appropriate Trust in AI-Assisted Decision-Making}




\author{Shuai Ma}
\orcid{0000-0002-7658-292X}
\affiliation{
  \institution{The Hong Kong University of Science and Technology}
  \city{Hong Kong}
  \country{China}
}
\email{shuai.ma@connect.ust.hk}

\author{Ying Lei}
\affiliation{
  \institution{East China Normal University}
  \city{Shanghai}
  \country{China}
}

\author{Xinru Wang}
\affiliation{
  \institution{Purdue University}
  \city{West Lafayette}
  \country{United States}
}

\author{Chengbo Zheng}
\affiliation{
  \institution{The Hong Kong University of Science and Technology}
  \city{Hong Kong}
  \country{China}
}

\author{Chuhan Shi}
\affiliation{
  \institution{The Hong Kong University of Science and Technology}
  \city{Hong Kong}
  \country{China}
}

\author{Ming Yin}
\affiliation{
  \institution{Purdue University}
  \city{West Lafayette}
  \country{United States}
}

\author{Xiaojuan Ma}
\affiliation{
  \institution{The Hong Kong University of Science and Technology}
  \city{Hong Kong}
  \country{China}
}

\renewcommand{\shortauthors}{Shuai Ma et al.}

\begin{abstract}

In AI-assisted decision-making, it is critical for human decision-makers to know when to trust AI and when to trust themselves. However, prior studies calibrated human trust only based on AI confidence indicating AI's correctness likelihood (CL) but ignored humans' CL, hindering optimal team decision-making. To mitigate this gap, we proposed to promote humans' appropriate trust based on the CL of both sides at a task-instance level. We first modeled humans' CL by approximating their decision-making models and computing their potential performance in similar instances. We demonstrated the feasibility and effectiveness of our model via two preliminary studies. Then, we proposed three CL exploitation strategies to calibrate users' trust explicitly/implicitly in the AI-assisted decision-making process. Results from a between-subjects experiment (N=293) showed that our CL exploitation strategies promoted more appropriate human trust in AI, compared with only using AI confidence. We further provided practical implications for more human-compatible AI-assisted decision-making.

\end{abstract}




\maketitle

\input{01-Introduction.tex}

\input{02-Related_Work.tex}
\input{03-Method_and_Pilot_Study.tex}
\input{04-Experiment.tex}
\input{05-Results.tex}

\input{06-Discussion.tex}

\bibliographystyle{ACM-Reference-Format}
\bibliography{main}


\end{document}

%% file: 01-Introduction.tex
\section{Introduction}



Artificial Intelligence (AI) systems are increasingly adopted in various decision-making scenarios \cite{dastin2018amazon, dilsizian2014artificial, khandani2010consumer, wang2014improving, yang2018insurance}. However, AI is still far from 100\% accurate in many real-world applications \cite{bansal2020optimizing, buccinca2021trust}. Besides, due to legal and ethical concerns, it remains risky for AI to make a decision autonomously, especially in high-stake domains such as medicine, criminal justice, etc. \cite{cai2019hello, lee2021human, binns2018s}. Hence, a paradigm named AI-assisted decision-making \cite{buccinca2021trust, zhang2020effect, wang2021explanations, bansal2021does} is proposed and widely studied in HCI and AI communities. In this paradigm, AI performs an assistive role by providing a recommendation, while the human decision-maker can choose to accept or reject AI's suggestion in the final decision.

\begin{figure*}[t]
	\centering 
	\includegraphics[width=\textwidth]{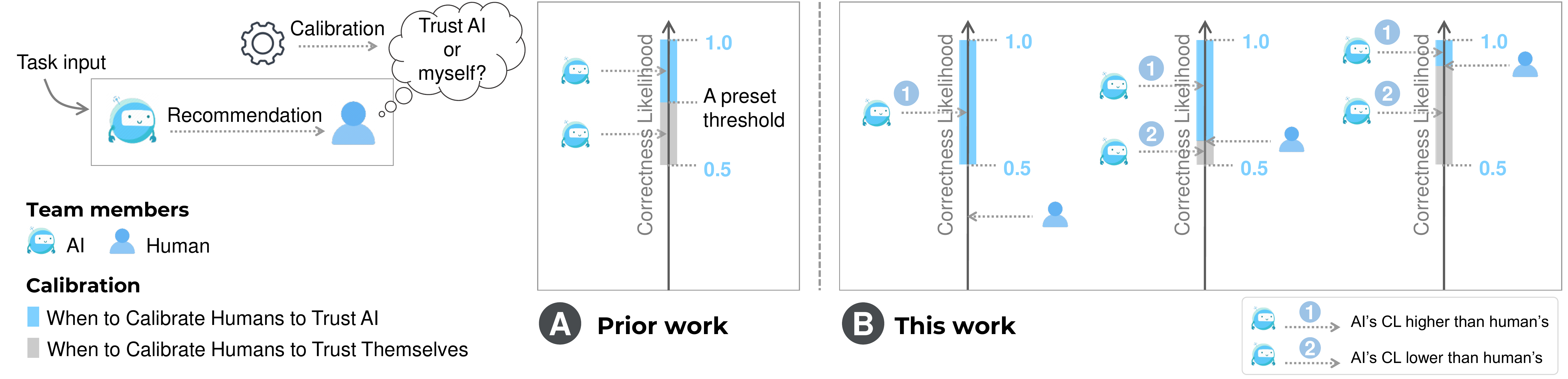}
	\caption{The difference between prior work and this work. (A) In prior work, AI's calibrated confidence is usually used to represent the AI's correctness likelihood (CL), a value ranging from 0.5 to 1.0. They calibrate humans' trust based on an empirically set threshold, i.e., when AI's confidence exceeds this threshold they will calibrate humans to trust the AI, and when AI's confidence falls below this threshold they will calibrate humans to distrust the AI (trust themselves). (B) In this work, besides considering AI's CL, we also estimate humans' CL in each task instance. We propose calibrating humans' trust based on the relative CL of both parties, rather than solely relying on whether AI's confidence is above a preset threshold. For example, if the AI's CL is higher than the human's, we will calibrate humans to trust AI; otherwise, we will calibrate humans to trust themselves.}
	\label{fig:paradigm}
\end{figure*}

One key challenge in AI-assisted decision-making is whether the human-AI team can achieve complementary performance, i.e., the collaborative decision outcome outperforming human or AI alone \cite{bansal2021does, lai2019human, zhang2020effect}. A critical step toward complementary performance is that human decision-makers could properly determine when to take the AI's suggestion into consideration and when to be skeptical about it \cite{zhang2020effect, buccinca2021trust, rastogi2020deciding}.
Since well-calibrated AI confidence scores can represent the model's actual correctness likelihood (CL) \cite{guo2017calibration, bansal2021does, bansal2019updates}, several recent studies propose different designs to help humans allocate appropriate trust to AI based on this information \cite{zhang2020effect, bansal2021does, rastogi2020deciding}.
For example, Zhang et al. \cite{zhang2020effect} directly display AI's confidence score to human decision-makers. Bansal et al. \cite{bansal2021does} show AI's explanations for the alternative predictions if the AI's confidence is below a threshold to make humans doubt the AI. Rastogi et al. \cite{rastogi2020deciding} propose leaving more time for humans to make a decision when the AI's confidence is lower than a threshold to reduce anchoring bias. 
Nevertheless, the empirical results from these studies are mixed at best \cite{zhang2020effect, bansal2021does, rastogi2020deciding, lai2022human}. There are two potential reasons. First, these works assume humans have an appropriate perception of their capability (CL) in a task instance to make reasonable decisions after knowing AI's CL. However, people usually have poorly-calibrated self-confidence that cannot reliably reflect their CL \cite{moore2020perfectly, miller2015meta, weber2004confidence, meyer2013physicians, kahneman2011thinking}. Second, these methods try to steer how much humans value AI's suggestions solely based on AI's correctness likelihood (illustrated in Figure \ref{fig:paradigm} (a)) while largely overlooking humans' correctness likelihood in each case. 
This poses a question: \emph{When AI's correctness likelihood is low (high) but that of humans is even lower (higher), should we still encourage humans to doubt (trust) the AI?}


To explore the answer to this question, in this paper, we propose a framework that aims to promote appropriate human trust in AI and complementary team performance according to the predicted human-AI correctness likelihood (CL) at a task instance level.
In this framework, as shown in Figure \ref{fig:paradigm} (b), we no longer have to calibrate human trust based solely on whether the AI's confidence exceeds a preset threshold. Instead, the CL of both humans and AI on a given task instance will be taken into consideration.
To verify the feasibility and efficacy of the proposed framework for promoting appropriate trust and complementary performance,
our investigation is divided into two phases: 1) How to model humans' capability (CL) on a given task? And 2) How to leverage human-AI capabilities (CL) to promote appropriate trust in AI-assisted decision-making? 

In the \emph{first} phase, based on the theories from cognitive science that humans usually adopt similar solutions to deal with similar problems \cite{cacciabue1992cosimo, goldstein2014cognitive, moyer1976mental, kahneman2011thinking}, we propose to estimate people's CL on a new task according to their performance in similar tasks. For example, if a person performs well on similar tasks, her CL on the current task is also likely to be high.
However, it is often difficult to obtain enough decision data to compute human performance on similar tasks. To solve this problem, we propose a method to first approximate a human's decision-making model (a mapping from task input to human decision), then apply this model to predict the human's possible decisions in similar task instances.
In the process, a question arises, \textbf{RQ1: How to effectively approximate a human's decision-making model?}
To explore the answer, we propose to combine data-driven initialization and interactive modification to derive the possible decision rules employed by each individual. And we design an interface called \emph{interactive rule set} for users to revise the initial model to better align with their inner decision-making process. We verified the appropriateness of the designed interface compared with another interface \emph{interactive decision tree} through a preliminary study (N=20). 
We take the system-initiated \& human-revised decision rule set as an approximated human decision-making model. For each new task case, we retrieve the closest cases from the existing task dataset (used for training the AI), then apply the derived models of individual decision-makers to get their likely predictions for those cases. Afterward, based on the estimated predictions and ground truth, we can calculate the probable performance of an individual, and further use this information to estimate the correct likelihood (CL) of that person on the current new task case. Combining the human CL and AI CL together, we can identify who has a higher capability in each task instance. Through a crowdsourcing study (N=30), we validated the effectiveness of our method in identifying complementary task instances (only one in the team can do it right) compared to the traditional AI confidence-based method.


In the \emph{second} phase of our work, after obtaining the estimated human-AI CL on an input task case, we further explore how to exploit this information to foster appropriate human trust and ultimately reach complementary performance in AI-assisted decision-making. In particular, we attempt to reduce human trust in AI when humans have a higher CL than AI, and increase human trust otherwise. Based on the relevant literature on people's cognitive processes \cite{bansal2019updates, nourani2021anchoring, buccinca2021trust, fogliato2022goes}, we propose three CL exploitation strategies to \emph{communicate} the CL of both sides to the responsible human decision-maker explicitly or implicitly, namely \emph{Direct Display}, \emph{Adaptive Workflow}, and \emph{Adaptive Recommendation}. Two related research questions emerge concerning these three CL exploitation strategies: \textbf{RQ2: How do different strategies affect human trust appropriateness and team performance?} And \textbf{RQ3: How do different strategies affect humans' perceptions and experiences in the decision-making process?}
Through a between-subjects crowdsourcing experiment with 293 participants, we found that our proposed three CL exploitation strategies resulted in more appropriate user trust in AI compared to baseline conditions, especially when the AI gave wrong recommendations. The three proposed CL exploitation strategies also led to improved team performance. However, different conditions did not lead to significantly different human perceptions or experiences in most subjective measures.

Our work provides a new perspective on promoting appropriate human trust in AI-assisted decision-making. In summary, our key contributions include:

\begin{itemize}
    \item We propose a framework to promote humans' appropriate trust in AI-assisted decision-making at a task instance level based on the capabilities of both sides.
    \item Accordingly, we design a method for estimating humans' CL on a new task instance with a data-driven initialization and interactive modification method to derive decision rules to estimate users' decision-making models.
    \item We conduct two preliminary studies to verify the appropriateness of the interactive decision rule creation interface, and to verify the effectiveness of the human CL modeling method.
    \item Based on the human-AI CL and related theories of humans' cognitive processes, we propose three CL exploitation strategies to foster humans' appropriate trust in AI explicitly or implicitly.
    \item We conduct a user study to analyze the impact of different CL exploitation strategies on user trust appropriateness, team performance, and user experience. Based on our key findings, we provide design implications for more effective human-AI collaborative decision-making.
\end{itemize}

%% file: 02-Related_Work.tex
\section{Related work}
\subsection{Trust Calibration in AI-Assisted Decision-Making}
Trust calibration refers to the correspondence between people's trust in the AI and the AI's actual capabilities \cite{lee1994trust}. When trust exceeds the AI's capabilities, over-trust leads to misuse, which refers to when people trust AI while they shouldn't \cite{lee2004trust, parasuraman1997humans}. Under-trust, when trust is less than the AI's capabilities, leads to disuse, which refers to people failing to use it when they should \cite{lee2004trust}. These flawed human-AI partnerships can result in costly and even catastrophic outcomes. Successful decision-making requires humans to calibrate their trust in AI on a case-by-case basis \cite{zhang2020effect, bansal2019beyond, bansal2019updates, bansal2021does, turnercalibrating}.

A pivotal approach to calibrating human trust is to convey AI's capability (also called reliability or trustworthiness) to humans \cite{bansal2021does, wang2021explanations, turnercalibrating, poursabzi2021manipulating}. There are several cues that can reflect AI's capability, such as the AI's accuracy (including stated accuracy \cite{yin2019understanding, rechkemmer2022confidence} and observed accuracy \cite{yin2019understanding, rechkemmer2022confidence, ma2022glancee}), explanation \cite{poursabzi2021manipulating, lai2020chicago, lai2019human}, the actual behavior/output \cite{gero2020mental, bansal2020optimizing, bansal2019beyond}, and confidence \cite{zhang2020effect, rastogi2020deciding, bansal2021does}, etc. For example, some works help people build a mental model of AI's error boundaries by observing AI's outputs \cite{bansal2019beyond}. Also, several studies expected that if humans were shown explanations for AI decisions \cite{bansal2021does, wang2021explanations, turnercalibrating, poursabzi2021manipulating}, they would be able to identify the trustworthiness behind the prediction.

One of the most commonly used capability indicators is AI's \emph{calibrated confidence score}, as well-calibrated confidence can accurately reflect the actual correctness likelihood (CL) of the AI in a specific task instance \cite{bansal2021does, bansal2019updates}. Therefore, many recent works calibrate human trust based on AI confidence. One line of work directly displays the calibrated confidence score to people. For example, Zhang et al. \cite{zhang2020effect} compared the effects of showing and not showing AI's confidence on people's trust calibration and task performance. Another line of work integrates AI confidence into the interface design. For example, Rastogi et al. \cite{rastogi2020deciding} discovered that if given longer thinking time, people would have more cognitive resources to invest in analytical thinking and reduce being anchored by AI. Therefore, they assigned different lengths of decision-making time to humans based on AI's confidence. In addition, Bansal et al. \cite{bansal2021does} developed an adaptive explanation strategy that explains the alternative predicted classes when the AI confidence is below a threshold, otherwise only explaining the top prediction.


There are two flaws in these works. On the one hand, they assume people have an appropriate perception of their capability (CL) in a task instance to make reasonable decisions after knowing AI's CL. However, people's subjective self-confidence usually cannot accurately represent their actual CL \cite{moore2020perfectly, miller2015meta, meyer2013physicians, kahneman2011thinking}. On the other hand, these approaches calibrate humans' trust only based on AI's CL and ignore human CL. For example, existing methods make people doubt AI when AI's confidence (CL) is low. But what if the human's CL is even lower? Note that the confidence of AI just represents a ``likelihood''; thus, a prediction with low confidence can still be correct, and a high-confidence prediction may also err. To solve these problems, our work proposes a novel method for calibrating human' trust based on human and AI capability (CL).

\subsection{Mental Model in Human-AI Collaboration}

Mental models are presentations of external reality that people use to interact with the world around them \cite{johnson1983mental, norman2014some}. In human-AI collaboration, some studies investigate building humans' mental model of the AI partner \cite{bansal2019beyond, gero2020mental, nourani2021anchoring, ma2022modeling}, so that humans know whether and when to assign a task to the AI. For example, Gero et al. \cite{gero2020mental} find those who win more often have better estimates of the AI agent's abilities in a cooperative game setting. Bansal et al. \cite{bansal2021does} help humans build a mental model for the AI system's error boundary, and they found that a good mental model can help humans achieve better performance. Besides building a mental model of how AI works, a faithful mental model of how human works is also essential. For example, in human-robot interaction, some works approximate human decision policy by modeling how people will behave in different environments \cite{ng2000algorithms, deng2018prediction}. However, little attention has been paid to leveraging the model of how humans make decisions in AI-assisted decision-making. In this paper, we approximate humans' decision-making (mental) models at the instance level (i.e., given a task instance, what prediction will people make), then based on the model, we can estimate humans' CL on a new task instance.

One approach to building humans' mental models is through data-driven methods. For example, in a loan approval task, Wang et al. \cite{wang2022will} construct a general human prediction model via a neural network with crowdsourcing data. Another approach is through rule-based methods. For instance, Bansal et al. \cite{bansal2019beyond} use simple rules to build humans' mental model of AI's error boundary, such as ``(age = old \& bloodePressure = high)''. Mozannar et al. \cite{mozannar2022teaching} ask humans to formalize their mental model of AI's error regions by writing a rule describing the region after solving a set of selected examples. Especially, rule-based methods have the advantage of interpretability \cite{kulesza2013too, lim2009and, lakkaraju2016interpretable, liao2021human, arrieta2020explainable}.
In this work, we propose to combine data-driven initialization and interactive rule modification to derive the possible decision-making mental model employed by individuals. This method has two advantages. First, it saves people's time by training an initial model via a small amount of user decision data, so that the model does not need to be built from scratch. Second, the model can also be presented to the user for manual interactive refinement.

\subsection{Cognitive Bias and Human Reliance in AI}
In human-AI interaction, as people are generally inclined to engage in System 1 thinking \cite{kahneman2011thinking}, there are often various cognitive biases, including common anchoring bias \cite{nourani2021anchoring}, confirmation bias \cite{nickerson1998confirmation}, automation bias/aversion \cite{cummings2004automation}, availability bias \cite{wang2019designing}, illusion of validity \cite{simkute2020experts}, etc. These cognitive biases can (negatively) affect people's trust in AI. For example, after observing model behaviors early on, people often have an anchoring bias towards AI's suggestions \cite{nourani2021anchoring}, leading to over-rely on AI's suggestions. People are also often brought by the illusion of validity of the information displayed by AI \cite{lai2019human, kaur2020interpreting, eiband2019impact}. For example, Kaur et al. \cite{kaur2020interpreting} find that the existence of explanations could mistakenly lead to data scientists' over-confidence that the model is ready for deployment. Eiband et al. \cite{eiband2019impact} find that even placebic explanations, which do not convey useful information, invoke a similar level of trust as real explanations do.

In order to reduce the adverse effects of cognitive biases on human-AI cooperation, existing works have proposed some mitigation methods. One way is to provide interventions to nudge people to engage deeper in System 2 thinking \cite{kahneman2011thinking}. For example, research on ``cognitive forcing'' has explored methods for pushing human decision makers to spend more time deliberating about problems \cite{buccinca2021trust, park2019slow, rastogi2020deciding}, such as asking humans to make independent predictions before seeing AI's suggestions \cite{buccinca2021trust} or employing a ``slow algorithm'' \cite{park2019slow}. These cognitive forcing functions are found to be able to \textbf{decrease} humans' AI reliance. Other mitigation methods include enabling people to actively explore the data \cite{wang2019designing, simkute2020experts}, explaining clearly and training users on how to use explanations/AI \cite{lai2020chicago}, giving arguments for non-predicted outcomes \cite{bussone2015role}, monitoring user's anchored status \cite{echterhoff2022ai}, showing prior probabilities of outcome \cite{wang2019designing}, etc.




In this work, we ``\emph{leverage}'' humans' cognitive biases to help us calibrate humans' trust by incorporating cognitive biases into adaptive interaction design. Specifically, we do not blindly increase or decrease people's trust in AI. Instead, we regulate the distribution of people's trust according to the CL of both parties. When AI's CL is higher, we utilize anchoring bias to promote people's trust in AI, and when human's CL is higher, we deploy cognitive forcing to promote human-independent analytical thinking.




%% file: 03-Method_and_Pilot_Study.tex
\section{Phase I: Modeling Humans' Correctness Likelihood on a Given Task Instance}

\subsection{Overall process of human correctness likelihood modeling}
\begin{figure*}[t]
	\centering 
	\includegraphics[width=1\linewidth]{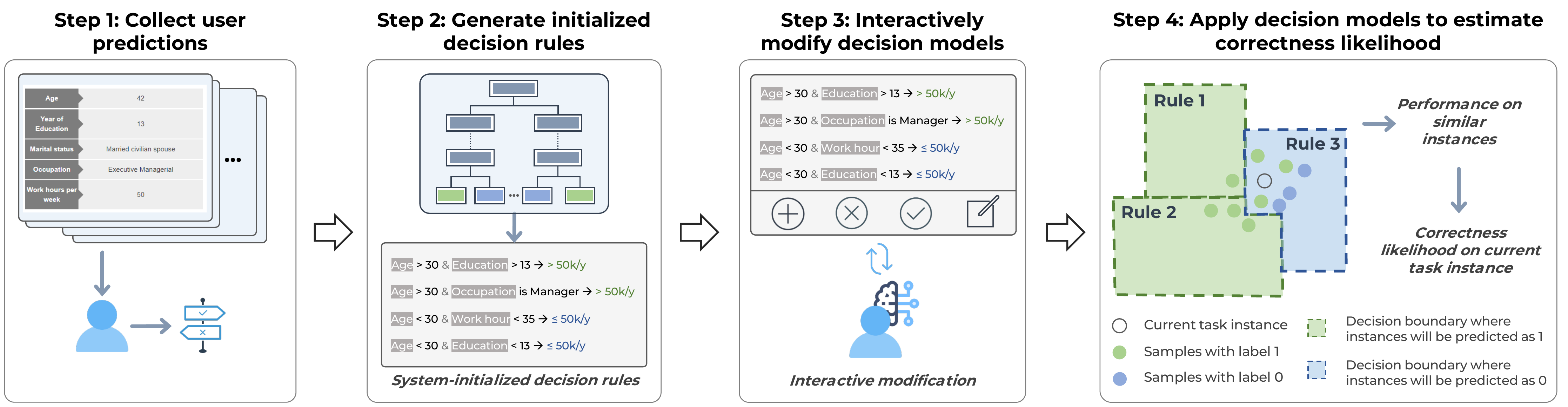}
	\caption{The human capability modeling process. The whole process goes through four steps. Step 1: Collect user predictions. Step 2: Generate initialized decision models. Step 3: Interactively modify decision models. Step 4: Apply decision models to estimate correctness likelihood.}
	\label{fig:modeling}
\end{figure*}

We investigate our human correctness likelihood (CL) modeling method in a typical AI-assisted decision-making scenario, where ground truth data is available for a task dataset but not for the current case. To promote more appropriate human trust in the AI-assisted decision-making process, the first phase of this work is to estimate humans' capabilities at a task instance level. Inspired by research in cognitive science which suggests that humans make decisions by weighing similar past experiences \cite{bornstein2017reminders, cacciabue1992cosimo, goldstein2014cognitive, moyer1976mental, kahneman2011thinking}, we propose to estimate humans' CL at a given task instance \cite{moore2020perfectly} based on their past performance in similar task instances. However, it is often difficult to obtain enough decision data to compute human performance in similar task instances, especially for a new task. 
To solve this problem, we design a method first to approximate a human's decision-making model (i.e., to get a mapping from task input to human decision). Then we apply this model to predict the human's possible decisions in similar task instances and calculate their potential accuracy in these instances compared to the ground truth. To answer the question, \textbf{RQ1: How to effectively approximate a human's decision-making model?}, we propose to combine data-driven initialization and interactive modification to derive the possible decision rules employed by each individual. 
Our proposed human capability modeling method goes through four steps, as illustrated in Figure \ref{fig:modeling}.

\begin{itemize}
    \item Step 1: Collect user predictions: we gather the decision data of people on a small number of sampled task instances.
    \item Step 2: Generate initialized decision models: we fit a classic decision tree model to infer humans' decision-making models and generate initial decision rules \cite{song2015decision, gennatas2020expert}.
    \item Step 3: Interactively modify decision models: due to the limited amount of training data, the initial model generated may not reflect people's actual decision-making model well. Therefore, we design an interactive interface for users to revise the initial model to make it better align with their inner decision-making process.
    \item Step 4: Apply decision models to estimate the correctness likelihood of new cases: We apply humans' approximated decision-making model to the neighbor cases of the current task case and compute humans' possible performance. Then, based on a distance-weighted method (see Eq. \ref{equation1}), we can estimate humans' CL in the current case.
\end{itemize}
In the following subsections, we present our task setup and introduce the details of the four steps through two small-scale studies (with Study I.1 focusing on Steps 1-3 and Study I.2 focusing on Step 4). 

\subsection{Task setup}
\label{task-setup}
\subsubsection{Task selection}
We chose \emph{income prediction} as our testbed which has been used in several previous studies on AI-assisted decision-making \cite{zhang2020effect, hase2020evaluating, ribeiro2018anchors, ghai2021explainable}. In this task, a participant was asked to predict whether a given person's annual income would exceed \$50K or not based on some demographic and job information. The data used for the task came from the Adult Income dataset \cite{datasetucl} in UCI Machine Learning
Repository. The entire dataset has 48,842 instances of surveyed individuals, each described by 14 attributes such as age, occupation, etc. These people's actual annual income was recorded and binarized (greater/less than 50K) as the ground truth for assessing participants' prediction accuracy. This task is suitable for our study since it requires little domain expertise and imposes relatively limited risks, and thus is amenable for non-expert participants \cite{ghai2021explainable}.

To ensure the task has a reasonable complexity for lay people to establish a decision-making model, following \cite{zhang2020effect, ghai2021explainable}, we selected the five most important features out of the 14 attributes as the final attributes presented to participants, determined by the feature importance values based on the feature permutation method \cite{altmann2010permutation}. These attributes include age, year of education, occupation, marital status, and work hours per week. This number of features is suggested to be appropriate for non-expert users to form a decision-making model by experiencing several task samples (e.g., Bansal et al. \cite{bansal2019beyond, bansal2019updates} established users' mental models of AI's error boundaries using a three-feature task). Future work can be extended to simulate humans' decision-making models in more complex tasks.

\subsubsection{AI model}
Same as \cite{ghai2021explainable}, we chose a logistic regression model (using a default setting from \emph{sklearn}) as our AI model to assist humans in making decisions in the selected income prediction task. As the logistic regression model directly optimizes Log loss, it can return well-calibrated confidence scores
\cite{platt1999probabilistic}.
Calibrated confidence of a model can provide an accurate probability of correctness for the model's predictions. For example, if a model makes a prediction on a sample with 0.6 confidence (calibrated), there will be a 60\% chance that the prediction is correct, or equivalently, if a model makes predictions on M samples with 0.6 confidence, there will be around 0.6 * M samples that are actually correctly predicted.
Note that some ML classifiers (such as SVM and neural networks) cannot directly generate calibrated confidence scores \cite{guo2017calibration, bansal2021does, zhang2020effect}, so post-hoc calibration is required (such as Platt Scaling or Isotonic
Regression \cite{platt1999probabilistic, guo2017calibration}).

Our model was trained based on a 70\% random split of the original dataset, while the prediction trials given to the participants in the experiment were drawn from the remaining 30\%. For any new task cases in the testing set, our human capability estimation method will retrieve similar cases from the training set to predict humans' CL.

\begin{figure*}[t]
	\centering 
	\includegraphics[width=1\linewidth]{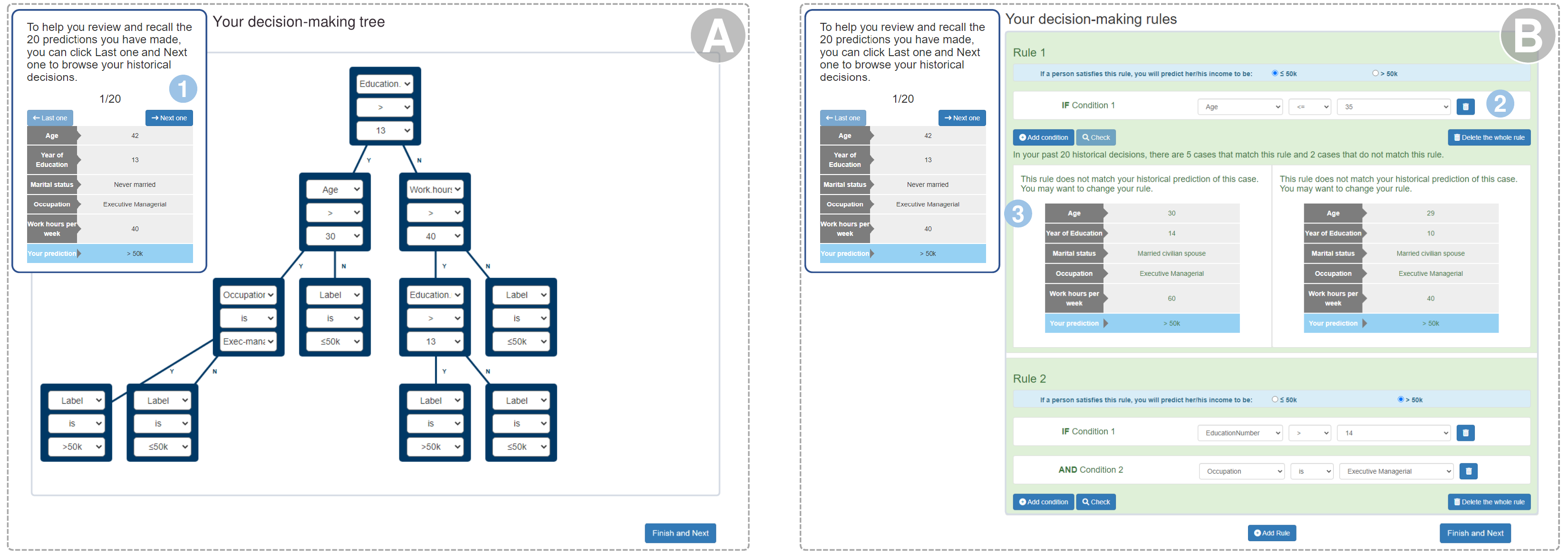}
	\caption{Two decision-making model creation interfaces. (A) The interactive decision tree. (B) The interactive rule set.}
	\label{fig:rule_interfaces}
\end{figure*}

\subsubsection{Task cases selection}
The selected task cases for the user studies satisfy several criteria. First, to make the human-AI teaming setting more suitable for pursuing complementary performance, humans' independent accuracy on these samples should be comparable to that of AI \cite{bansal2021does, zhang2020effect}. Second, these cases should follow the data distributions in the test set \cite{wang2021explanations}. Third, AI's confidence scores in these samples should be well-calibrated to reflect its actual CL \cite{wang2021explanations, zhang2020effect}.

To keep the user studies at a proper length without causing fatigue in participants, we selected 40 task cases, which are split into two batches. The first 20 samples are used to get humans' decision data and build their decision-making models computationally. The remaining 20 samples are used in the main AI-assisted decision-making task. While the two batches of samples are fixed for all participants, the presentation order of samples inside each batch was randomized. 
To make AI performance comparable to humans' independent accuracy, following \cite{bansal2021does}, we first conducted an additional pilot study to determine the average prediction accuracy of unassisted humans over 20 randomly picked task instances, which was around 70\% according to the results. We then selected 40 task samples over which the AI model had a 70\% accuracy with equal positive and negative labels, as well as equal false positive and false positive rates (similar setting as \cite{bansal2021does}).
To guarantee the representativeness of the selected samples, we made sure that most of the common values of each feature were included in these 40 samples. We also carefully controlled the AI's confidence in these instances to make it align with AI's actual CL. 
Specifically, out of the 20 samples in each batch, half of them had a confidence score lower than 0.7 (representing low-confidence samples, with an average value of 0.6), of which 6 samples were correctly predicted by the AI (the CL was $6/10 = 60\%$). Another half of them had a confidence score higher than 0.7 (representing high-confidence samples, with an average of 0.8), of which 8 samples were correctly predicted by the AI (the CL was $8/10 = 80\%$). 


Once the AI models and task cases are ready, we conducted a lab study to explore the suitable interface for non-expert users to interactively revise their decision-making models. Note that one may ask, \emph{why do we need to model human correctness likelihood (CL)? Just as AI's confidence can indicate its CL, can people's self-confidence represent their CL?} We carried out a small-scale user study and found that participants had poorly-calibrated subjective confidence. That is, the correlation between their actual accuracy and self-reported confidence is statistically unrelated, suggesting that self-reported confidence is not a reliable human CL indicator. The details can be found in the supplementary material.

\subsection{Study I.1: Comparison of interfaces for users to specify their decision-making models}
\label{decision_rule}
According to existing research, rules are considered to be an appropriate mechanism for approximating human decision-making processes \cite{gennatas2020expert, furnkranz2012foundations, furnkranz2020cognitive}. Humans, on the other hand, often make judgments based on decision tree-like structures \cite{damez2005fuzzy}. Hence, to compare the efficacy of these two representations, we design two interfaces for displaying and interactively updating humans' decision-making models (i.e. \emph{interactive decision tree} versus \emph{interactive rule set}). Both interfaces share the same initialization method where we fit a decision tree model (default setting from \emph{sklearn}) on the human decision data from the first 20 task cases. We chose the decision tree model instead of a black-box model because it can be easily understood even by people without machine learning knowledge \cite{damez2005fuzzy}.

These two interfaces are shown in Figure \ref{fig:rule_interfaces}. The \emph{interactive decision tree} interface (Figure \ref{fig:rule_interfaces} (a)) directly displays the decision tree model generated in the backend. On the interface, humans are first shown a tutorial about how to interpret and modify the decision tree (not included in the figure). Then, they can browse their past decision data on the first 20 instances to recall their decision-making rationale (Figure \ref{fig:rule_interfaces} (1)). Finally, they can add, delete or modify any tree node to reflect their actual decision process.
The \emph{interactive rule set} interface (Figure \ref{fig:rule_interfaces} (b)) presents the set of \emph{if-then} rules converted from the decision-tree \cite{lim2009and}. 
With this interface, similarly, humans first view a tutorial, next revisit their historical decision data, and finally they can add, delete, or modify any rules or conditions in a rule (Figure \ref{fig:rule_interfaces} (2)). 

\subsubsection{Study procedure and participants.} We conducted a between-subjects study, recruiting 20 participants (8 Female, 12 Male, average age: 27) from a local research university to build their decision models using the assigned interface (10 for each condition). 
After giving their consent, they followed a tutorial to familiarize themselves with the income prediction task. Then, they proceeded to finish 20 prediction tasks (the first batch) without the help of AI. Upon completion, they were asked to use the assigned interface to create their decision model. We mainly focused on their qualitative perceptions of the interface, so we carried out an exit interview with them at the end of the study.

\subsubsection{Results} According to the interview results, seven out of the 10 participants using the \emph{interactive decision tree} interface reported that their actual decision process could not be well represented by a decision tree. For example, P3 (Male, 30, little knowledge in AI) noted, ``\emph{My actual decision process was not a single (decision) tree. Sometimes, I use `age' as the first criterion, but sometimes, I use the `year of education' as the main factor. }'' Furthermore, three out of the 10 participants found the decision tree to be visually complex. For instance, P9 (Male, 26, little knowledge in AI) mentioned that ``\emph{The tree is hard for me to read in a short time.}'' In comparison, the \emph{interactive rule set} interface was considered to be more visually interpretable and more in line with participants' decision-making process by users in this condition. Therefore, in the final version, we employ the \emph{interactive rule set} interface for participants to revise their decision-making models interactively.

Based on participants' feedback, we also improved the \emph{interactive rule set} interface. For each rule, we provide a ``check'' button, clicking on which allows users to check how many of their historical decisions conflict with this rule and whether this rule conflicts with other created rules (Figure \ref{fig:rule_interfaces} (3)).


\subsubsection{Discussion}
While a decision rule set is better suited for simulating human decision-making models, it also has some limitations. First, there are sometimes edge cases that are difficult to cover by limited decision rules \cite{furnkranz2020cognitive}. For these cases, we now use the system-initialized model to cover. Second, some users make decisions based on intuition, which can not be formulated as an explicit set of rules. Third, it may be difficult for non-expert users to form accurate decision rules by experiencing only a small set of task samples. We will discuss these in more detail in Sec. \ref{limitation}.

\subsection{Study I.2: Performance testing of our human correctness likelihood estimation method}
Based on the user-revised decision-making model, we can get their possible predictions for $N$ similar task instances retrieved from the training set. And comparing their possible predictions and ground-truth labels (already known), we can compute humans' potential performance on these task instances to obtain an estimated CL for the current task instance. We empirically set the number $N$ to 10 in this work to achieve a trade-off between sufficient similarity and coverage. If the number is set too large, a lot of dissimilar samples will be calculated and if the number is set too small, the sample size is insufficient to obtain a stable accuracy value. Note that the number can be different in other tasks with different properties. We calculate human correctness likelihood $CL_c$ on the current task instance $I_c$ based on the following equation.

\begin{small}
\begin{equation}
    CL_c = \frac{\sum_{i=1}^{N} w^i \cdot IF(\hat{y}^{i} = y^i, 1, 0) + (1-w^i) \cdot 50\%}{N}, \quad where\ w^i = \frac{\alpha}{\alpha + d(\mathbf{x}_c,\mathbf{x}_n^i)}.
\label{equation1}
\end{equation}
\end{small}

where $\hat{y}^{i}$ is the human possible prediction in the $i$-th neighbor instance $I_n^i$, and $y^i$ is the ground-truth label of that instance. $IF(\hat{y}^{i} = y^i, 1, 0)$ means if $\hat{y}^{i} = y^i$, returns 1, otherwise, returns 0. And $w^i$ is the weight of each neighbor instance, $d(\mathbf{x}_c,\mathbf{x}_n^i)$ is the Euclidean distance between the current task instance $I_c$'s feature vector $\mathbf{x}_c$ and its neighbor instance $I_n^i$'s feature vector $\mathbf{x}_n^i$. We can see that the weight is negatively correlated with the distance. More similar neighbor instances will have a greater impact on performance computations. For example, if a human can make a correct prediction for a very close neighbor instance ($d(\mathbf{x}_c,\mathbf{x}_n^i) \rightarrow 0$), it will contribute $1/N$ to CL. If a human makes a correct prediction for an (extremely) distant task instance ($d(\mathbf{x}_c,\mathbf{x}_n^i) \rightarrow \infty$), the distance factor will discount its contribution and move $w^i$ closer to 0 (i.e., it only contributes $0.5/N$ to CL, which is equal to random guessing). We set the parameter $\alpha$ to 2 based on the median Euclidean distance between any two instances in the training set. While other values may be more appropriate, we leave this to future work.

Combining the human CL and AI CL (calibrated AI confidence), for a new task instance, we can estimate which member in the human-AI team has a higher correctness likelihood. Next, we verify the effectiveness of our method with two objectives. First, the estimated human CL should be significantly correlated with the actual human accuracy. Second, recall that a key purpose of our approach to modeling human capabilities is to better distinguish when to trust the AI and when to trust themselves. So we focus on the \emph{complementary region}, where for each case, only one member of the human-AI team can make a correct prediction. If the human is estimated to have a higher CL on a case, this case will be labeled ``human better''; otherwise, ``AI better''. In comparison, in the AI confidence-based method, same as previous works \cite{zhang2020effect}, when the AI's confidence exceeds the set threshold (0.7), we regard this case as ``AI better'' and otherwise ``human better''. We quantify the effectiveness as the $recall$ of complementary cases, i.e., the ratio of complementary cases that are correctly predicted by our method out of the whole \emph{complementary region}. We didn't focus on the $precision$ because in the case where both humans and AI can make correct or incorrect predictions, whoever has a higher likelihood won't lead to significantly different consequences.

\subsubsection{Study procedure and participants.} In the same setting as Study I.1 (Sec. \ref{decision_rule}), we conducted a crowdsourcing study to compare the effectiveness of our method and the AI confidence-based method. We recruited 30 participants from Prolific\footnote{www.prolific.co\label{prolific}} (18 Female, 11 Male, 1 non-binary, aged from 21 to 61, avg 35, all reside in the US). The study procedure was the same, except that we also asked participants to complete the remaining 20 tasks after creating their decision rules (again, without the assistance of AI, so that we could measure humans' independent correctness).

\subsubsection{Results.} We found that based on the auto-generated human decision-making model, the prediction accuracy of participants' decisions on the last 20 task instances was 77.5\%. In comparison, based on the human-revised decision-making model, the accuracy was 80.7\%. This shows a slight but not significant improvement. We speculate that this is because the default decision tree model is already close to the human decision-making process, so participants can only make minor adjustments to the initialized rules.
Following \cite{depaulo1997accuracy, rechkemmer2022confidence}, we calculated the Pearson correlation between our estimated humans' average CL and their actual accuracy on the last 20 tasks. The result showed a significantly positive correlation ($r$=0.482, $p$<.01).
Furthermore, we found that our human-AI CL method could recall 76.4\% of the \emph{complementary region} on average, while the AI confidence-based method could recall 66.7\% of the \emph{complementary region} on average. Paired T-tests showed significant differences ($p$<.05). The results validated that our method was more effective than the traditional AI confidence-only methods at guessing the human-AI CL on complementary cases.

\subsubsection{Discussion}
We note that our method highly relies on the accuracy of the approximated human decision-making models. Besides, although our method is better than the AI confidence-based method in identifying complementary cases, the improvement is not very large. We speculate it might be due to the limited complementarity of humans and AI, which affects the superiority of our method. We will discuss this issue in the final Discussion.



%% file: 04-Experiment.tex
\section{Phase II: Communicating Human's and AI's Correctness Likelihood to Promote Appropriate Trust} 
The second phase of this work is to explore how to integrate the modeled human-AI correctness likelihood (CL) to empower the AI-assisted decision-making process. Specifically, we propose three different strategies to exploit CL, i.e., \emph{Direct Display}, \emph{Adaptive Workflow}, and \emph{Adaptive Recommendation}. Then, through a between-subjects experiment, we aim to investigate two research questions: \textbf{RQ2: How do different CL exploitation strategies affect human trust appropriateness and team performance?} and \textbf{RQ3: How do different CL exploitation strategies affect humans' perceptions and user experiences in the decision process?}


\subsection{Experimental Conditions and Interface Design}
To help people realize when to refer to the AI's suggestion and when to rely on themselves, one intuitive mechanism is to \emph{explicitly} display the human and AI CL information to human decision-makers.

\begin{figure*}[h]
	\centering 
	\includegraphics[width=\linewidth]{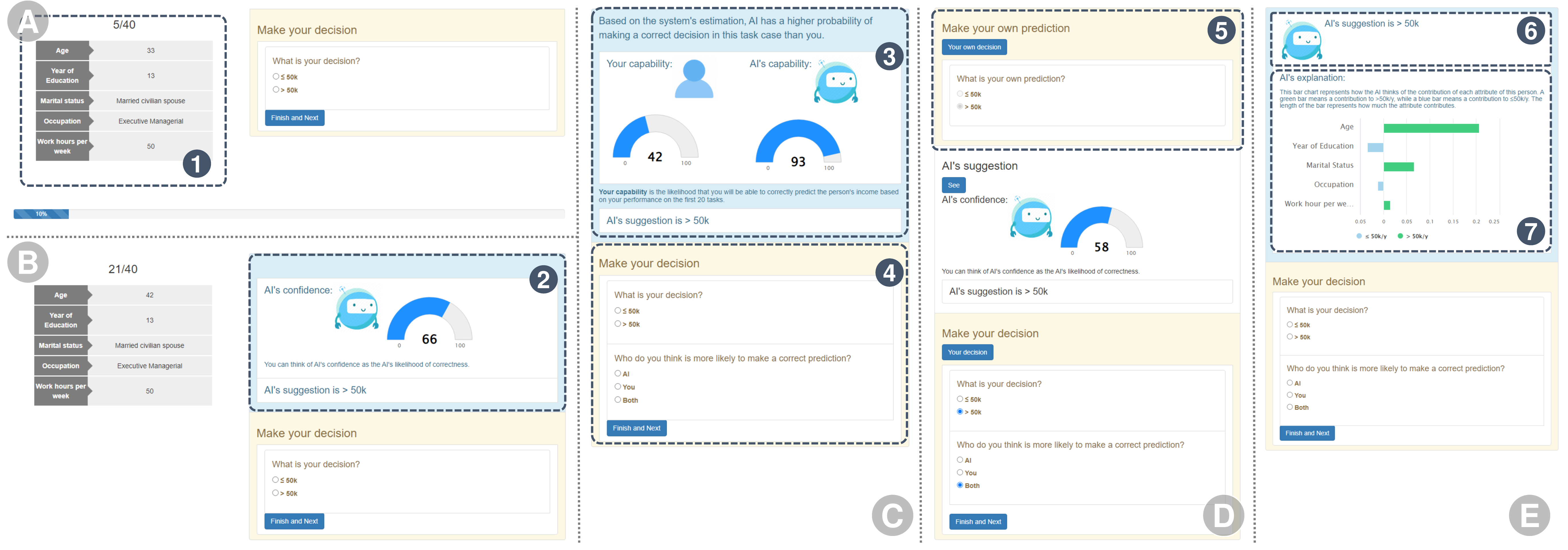}
	\caption{Interface of different conditions. (A) \textbf{Human Only}. (B) \textbf{AI Confidence}. (C) \textbf{Direct Display}. (D) \textbf{Adaptive Workflow}. (E) \textbf{Adaptive Recommendation}. The interfaces of all conditions share a similar layout: the left side is a person's profile area and the right side is a decision-making area. To save space, we do not draw the person profile area repeatedly in Figure \ref{fig:condition-interfaces} (C), (D), (E).}
	\label{fig:condition-interfaces}
\end{figure*}

\begin{itemize}
    \item \textbf{Direct Display}: We directly present the estimated human and AI CL and the AI's recommendations to the human (Figure \ref{fig:condition-interfaces} C) in this condition. To be more specific, on the experimental website, alongside the profile area (the five attributes of the person to predict, Figure \ref{fig:condition-interfaces} A1), the system illustrates the estimated CL of the human and AI side by side (Figure \ref{fig:condition-interfaces} C3). At the top of this area is a summary sentence, "According to the system's estimation, in this task case, the AI (you) might have a higher probability of making a correct decision than you (the AI)". Below are two gauge graphs showing the CL values of humans and AI, respectively, followed by the recommendation (i.e., the predicted income) from AI. Finally, people need to input their final decision (Figure \ref{fig:condition-interfaces} B4).
\end{itemize}

In this condition, it is up to human users to decide how to interpret the CL information and whether to trust the AI. 
However, we acknowledge that the estimates of humans' and AI's capabilities are far from perfect, and relying on this information to assess AI's suggestions may have serious consequences, especially in high-risk areas. For example, in a clinical decision-making scenario, physicians may develop false self-confidence in their diagnosis if our model overestimates their abilities. 
To mitigate this issue, we propose two other \emph{implicit} CL exploitation strategies based on theories in cognitive science. On the one hand, according to the \emph{anchoring bias} theory in decision-making \cite{epley2006anchoring, furnham2011literature, rastogi2020deciding, buccinca2021trust, epley2001putting}, if human decision-makers have access to anchors (such as AI's opinions), they are likely to diminish further exploration of alternative hypotheses and \textbf{increase} humans' reliance on AI. On the other hand, research on ``cognitive forcing'' has explored methods for pushing human decision-makers to spend more time with deliberating about problems \cite{buccinca2021trust, park2019slow, rastogi2020deciding}.
These cognitive forcing functions are found to be able to \textbf{decrease} humans' reliance on AI. Based on these theoretical supports, we propose the following condition.




\begin{itemize}
    \item \textbf{Adaptive Workflow}: In this condition (Figure \ref{fig:condition-interfaces} D), we adaptively change the order of human and AI decisions based on the estimated human and AI CL. If the predicted human CL is higher than that of the AI, our interface will first ask human users to input their initial decision and then reveal the AI's recommendation (Figure \ref{fig:condition-interfaces} D5). Likewise, if the AI's CL is estimated to be higher than the human's, our interface will directly present the AI's suggestion to the human (Figure \ref{fig:condition-interfaces} D5 will not be displayed). Both cases require the human to make the final decision after reviewing the AI's recommendations. 
\end{itemize}

Another implicit way to leverage \emph{cognitive forcing} to promote appropriate trust is not to show AI suggestions to people if they have higher CL than the AI. However, this will prevent people from taking advantage of AI's assistance in such cases. A trade-off solution is to provide AI's explanations but not the prediction result to users when humans have a higher CL than AI so that they have to make their own decisions. Garhos et al. \cite{gajos2022people} found that people engaged more in analytical thinking based on their own knowledge if AI's explanations were shown without concrete recommendations. Therefore, we propose the following condition.

\begin{itemize}
    \item \textbf{Adaptive Recommendation}: In this condition, the AI shows the explanation of its prediction (generated by LIME \cite{ribeiro2016should}, a widely used XAI method) by default. We control the display of AI's recommendation based on the comparison between the estimated human and AI CL. If the AI's CL is higher, our interface will display AI's recommendation (Figure \ref{fig:condition-interfaces} E6) along with its explanation (Figure \ref{fig:condition-interfaces} E7). If the human CL is higher, our interface will \emph{not} disclose the AI's recommendation to users (Figure \ref{fig:condition-interfaces} E6 hidden). 
\end{itemize}

Besides the above-mentioned three conditions, we also include two baseline conditions following \cite{bansal2021does, rastogi2020deciding}.
\begin{itemize}
    \item \textbf{Human Only}: In this condition, humans must make their own decisions without any AI assistance (Figure \ref{fig:condition-interfaces} A).
    \item \textbf{AI Confidence}: In this condition, humans are presented with AI's recommendation and its calibrated confidence but \emph{without} human CL information (Figure \ref{fig:condition-interfaces} B2). We incorporate this baseline because it is a broadly acknowledged design to calibrate humans' trust in AI-assisted decision-making \cite{zhang2020effect, bansal2021does}.
\end{itemize}

The interfaces were tested through a pilot study to ensure that the workflow was clear for participants to follow.

\subsection{Study Design}
We employ a between-subjects study design with the five conditions. The study is approved by the University IRB.

\subsubsection{Task and Procedure}

\begin{figure*}[h]
	\centering 
	\includegraphics[width=\linewidth]{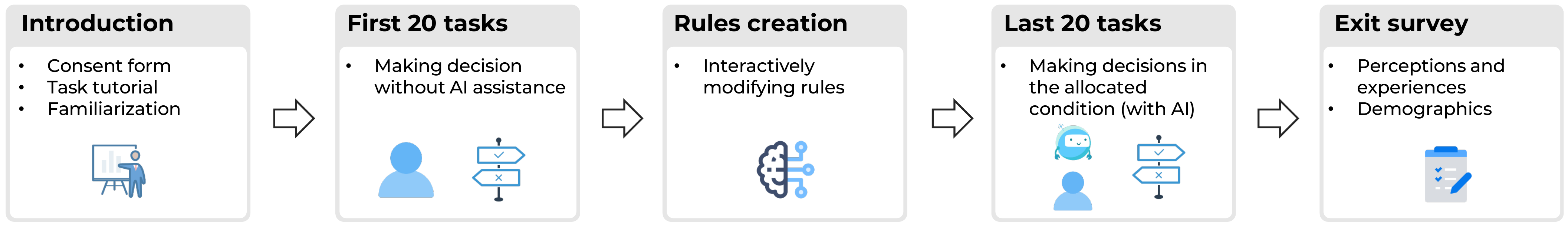}
	\caption{Procedure of the experiment. Participants go through five stages in the whole study.}
	\label{fig: procedure}
\end{figure*}

We adopted the same task as in Phase I (Sec. \ref{task-setup}), i.e., predicting whether a person's annual income exceeds \$50K. Participants went through five stages during the study (shown in Figure \ref{fig: procedure}): (1) Introduction: After obtaining the consent of participants, we provided a tutorial walk-through to familiarize them with the task where we detailed the meaning and value range of each attribute in the profile table to the participants. For each attribute, we presented a graph showing the distribution of the corresponding income in the entire dataset, giving participants a basic understanding of the salary situation. We inserted two attention-check questions at the end of the tutorial to help filter out participants who did not read the introduction carefully. After the tutorial, we provided participants with two training examples with ground truth. (2) First batch of 20 tasks: Next, participants proceeded to complete the first 20 task cases independently (no AI advice or ground truth information was displayed). (3) Interactive decision rule creation: Participants entered the decision rule creation page (see Sec. \ref{decision_rule} for details). (4) Second batch of 20 tasks: Once done customizing their own decision rules, participants moved on to the last 20 task cases, this time, with AI's assistance (except for \emph{Human Only} condition). Depending on the assigned conditions, different interfaces were presented to the participants (as shown in Figure \ref{fig:condition-interfaces}). (5) Exit survey: Finally, participants were asked to fill out an exit survey in which we collected basic demographic information as well as subjective measures and open-ended feedback about their perceptions in the decision-making process, which are described later in Sec. \ref{questions}.

\subsubsection{Participants}


We recruited 300 participants (60 for each condition) from Prolific\textsuperscript{\ref{prolific}}. To ensure high-quality responses, all participants met the following criteria: (1) residing in the United States (as the task was to predict income for adults in the United States); (2) at least 99\% approval rate for previous submissions; (3) using English as the first language; (4) owning a bachelor's degree or above; and (5) using a desktop computer for the experiment. The study followed a between-subjects design, so we did not allow any repeated participation. In total, we got 293 complete submissions. After filtering based on the attention-check questions, we obtain 289 valid responses (\emph{Human Only}: 59, \emph{AI Confidence}: 59, \emph{Direct Display}: 59, \emph{Adaptive Workflow}: 56, \emph{Adaptive Recommendation}: 56). Among the final participants, there were 174 self-reported male, 110 female, and 5 non-binary. A total of 77 participants were aged 18-29, 116 aged 30-39, 45 aged 40-49, 28 aged 50-59, and 23 aged over 59. Participants also rated their knowledge of artificial intelligence: 40 had no knowledge, 205 knew basic concepts in AI, 43 had used AI algorithms, and one was an expert in AI. To motivate high-quality work, in addition to the base payment, we gave participants a \$0.50 bonus if their overall accuracy exceeded 80\%. The entire study lasted about 20 minutes. The average wage for participants was about \$9.34 per hour.

\subsection{Evaluation Metrics}
\label{questions}

\subsubsection{Measures for RQ2.} We investigate the effects of different conditions on humans' trust appropriateness and human-AI team performance through two main measurements. (1) \emph{Human-AI Agreement} \cite{zhang2020effect, bansal2021does}: the fraction of tasks where the participant's final decision agreed with the AI's recommendation, whether it is right or wrong. (2) \emph{Team Performance} \cite{bansal2021does, zhang2020effect, rastogi2020deciding, wang2021explanations}: the final decision accuracy. We also collected participants' \emph{Perceived correctness likelihood (CL)}, where in each task instance, we asked participants which one (human, AI, or both) they thought had a higher CL.
\subsubsection{Measures for RQ3.} Here, we focus on participants' experiences and perceptions in different conditions. Specifically, referring to and adapted from related works, we investigate the following subjective measures as 7-point Likert scale questions in the exit survey (1: Strongly Disagree, 7: Strongly Agree): (1) \emph{Trust in AI} \cite{ghai2021explainable, buccinca2021trust}; (2) \emph{Confidence in the decision-making process} \cite{lai2022human, rechkemmer2022confidence}; (3) \emph{Perceived complexity of the system} \cite{buccinca2021trust}; (4) \emph{Mental demand} \cite{hart2006nasa, lai2022human, ghai2021explainable, buccinca2021trust}; (5) \emph{Perceived autonomy} \cite{hong2019racism}; (6) \emph{Satisfaction} \cite{ghai2021explainable}; (7) \emph{Future use} \cite{brooke1996sus}; (8) \emph{Trust in the estimation of human-AI CL}; (9) \emph{Perceived usefulness of estimation of human-AI CL} \cite{laugwitz2008construction}; (10) \emph{Perceived helpfulness to decide when to trust the AI} \cite{laugwitz2008construction}); and (11) \emph{Acceptance of estimation of their CL}. Besides these questions, we also asked participants open-ended questions about how they used and perceived the communicated human-AI CL, and how their decision-making processes were affected by different interface designs. Detailed questions can be found in the supplementary material.

\subsubsection{Analysis Methods}
\label{analysismethod}

We conducted mixed-methods analyses on the aforementioned metrics. For quantitative analysis of the objective data for RQ2 and participants' subjective data for RQ3, since most of the data did not follow a normal distribution, we carried on non-parameter tests. Specifically, for pair-wise comparison, we ran Mann-Whitney U Test or Wilcoxon Signed Ranks Test based on whether the sample was from the same group of participants. And for analysis among more than two groups of participants, we ran Kruskal-Wallis Test and post-hoc analysis with Bonferroni correction. For qualitative analysis, two authors coded the open-ended feedback via inductive thematic analysis \cite{hsieh2005three}. The final themes were discussed and harmonized over several iterations, and specific examples were identified from the source texts for demonstration in this paper.


%% file: 05-Results.tex
\section{Results}



\subsection{Effects of CL Exploitation Strategies on Team Performance and Human Trust in AI}
We organize the results into two parts. In the first part, we analyze the participants' overall team performance and trust in AI. In the second part, we dig deeper into the data and analyze the results according to different situations (e.g., different human-AI CL). All results are organized into ``Findings'' for easy reading.

\textbf{Part 1}

\begin{figure*}[t]
	\centering 
	\includegraphics[width=0.5\linewidth]{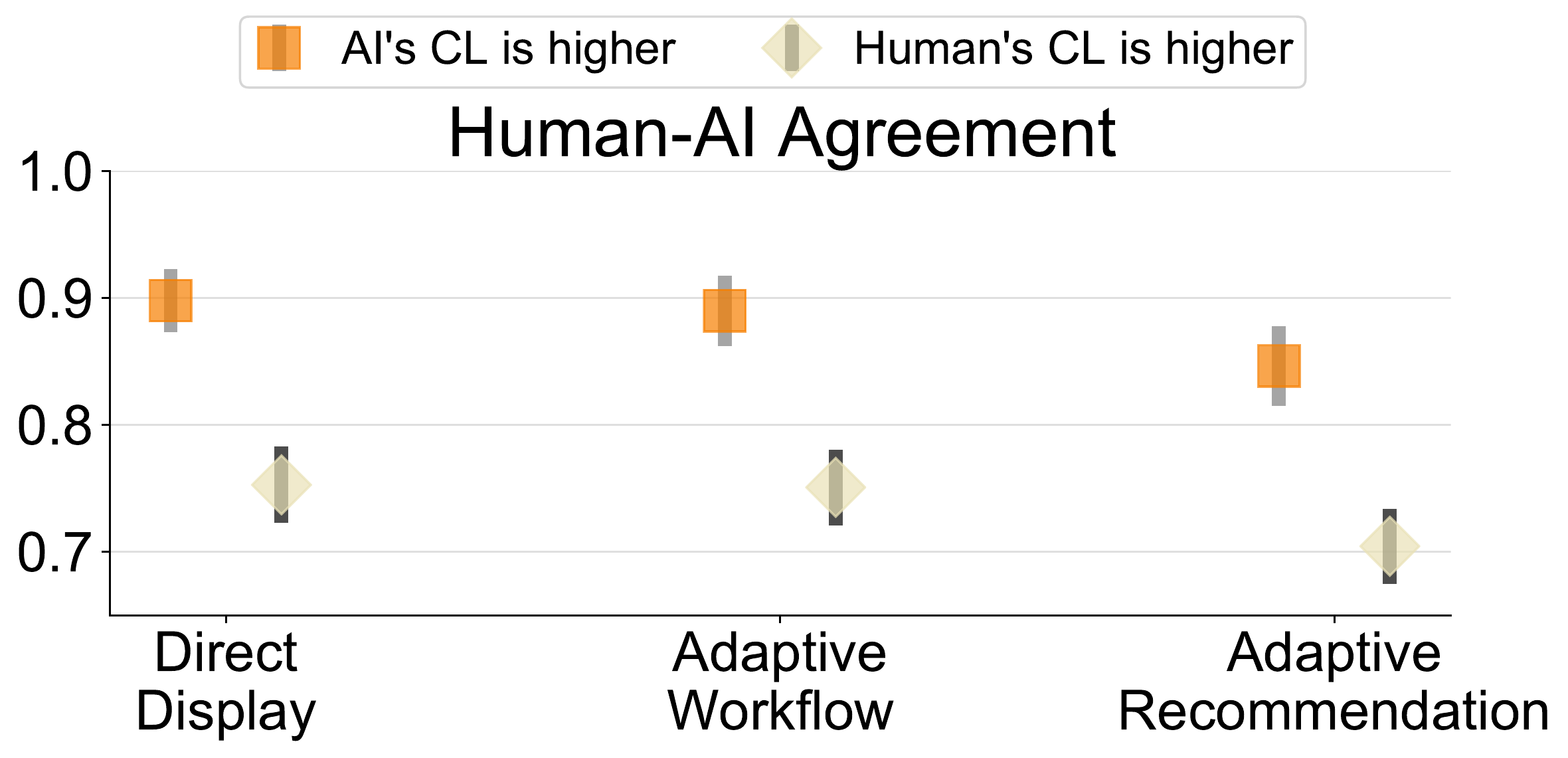}
	\caption{Human-AI agreement in different human and AI correctness likelihood (with mean and 95\% confidence interval). We can see that when the AI's CL is higher than the human's, participants tend to agree with AI more in the final decision.}
	\label{fig:agreement}
\end{figure*}

\textbf{Finding 1: The trend of human trust in AI was consistent with the trend of estimated human-AI CL in the proposed three conditions.}
Since the basic intention of our three designs is to make people rely more on the member with higher CL in the human-AI team, we want to see if our approaches made people trust AI \textbf{\emph{more}} when the AI's CL was higher and trust AI \textbf{\emph{less}} otherwise.
Figure \ref{fig:agreement} shows the human-AI agreement in different human-AI CL under three conditions. Results showed that all three conditions made people's agreement with AI significantly higher when AI's CL was higher than that when the human's CL was higher ($p$<.001 in all three conditions).


\begin{figure*}[h]
\centering
\subfigure[]{
\begin{minipage}[t]{0.5\linewidth}
\centering
\includegraphics[width=\linewidth]{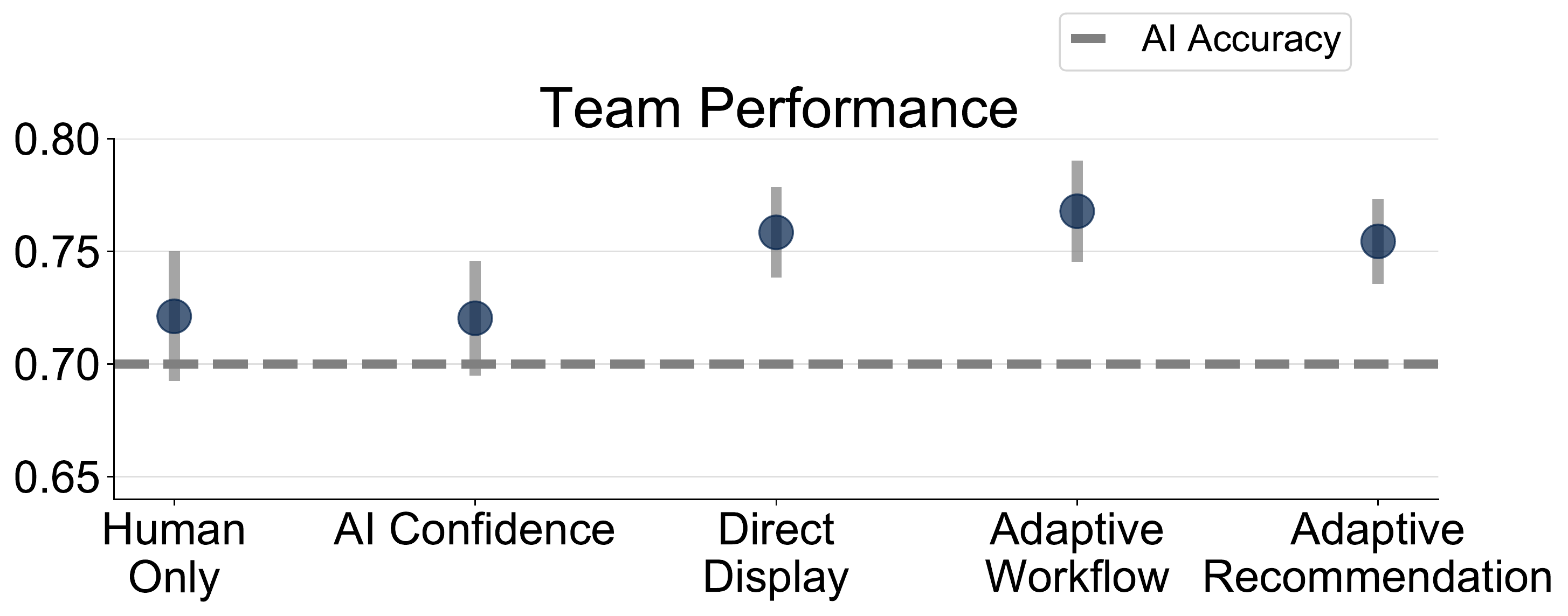}
\end{minipage}%
}%
\subfigure[]{
\begin{minipage}[t]{0.5\linewidth}
\centering
\includegraphics[width=\linewidth]{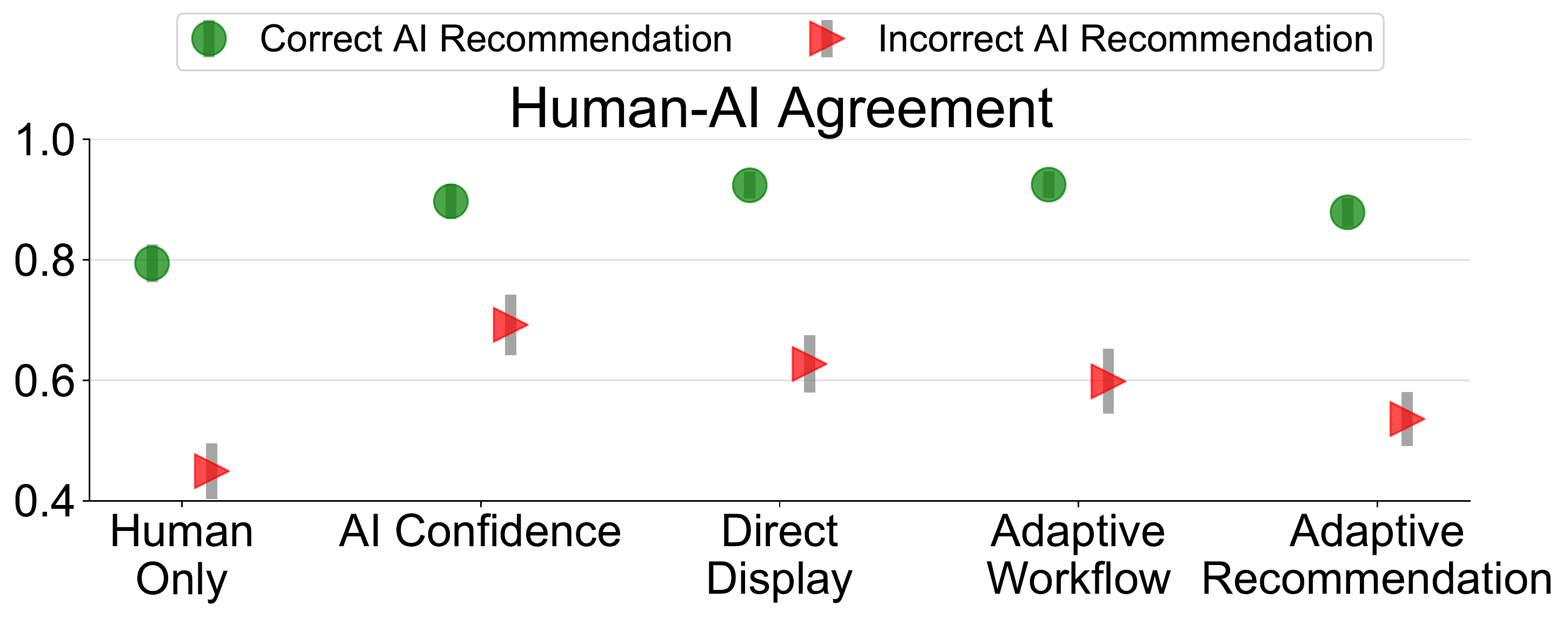}
\end{minipage}%
}%
\centering
	\caption{Overall team performance and trust appropriateness (with average accuracy and 95\% confidence interval) in different conditions. (A) The overall team performance in five conditions. The proposed three communication strategies achieve complementary performance compared with AI accuracy (0.7) and \emph{Human Only} (0.72). (B) Humans' trust appropriateness which is indicated by human-AI agreement when AI gives correct recommendations and when AI gives wrong recommendations.}
	\label{fig:team_performance}   
\end{figure*}

\textbf{Finding 2: The proposed three CL exploitation conditions achieved complementary performance while the \emph{AI Confidence} condition did not.} Figure \ref{fig:team_performance} (a) shows the overall team performance (i.e., the accuracy of humans' final decisions) in all conditions. The team performance in \emph{Direct Display} ($M$=0.758, $SD$=0.078), \emph{Adaptive Workflow} ($M$=0.767, $SD$=0.085), \emph{Adaptive Recommendation} ($M$=0.754, $SD$=0.072) surpassed both AI alone (0.7) and \emph{Human Only} ($M$=0.721, $SD$=0.113). However, the team performance of \emph{AI Confidence} ($M$=0.720, $SD$=0.099) did not outperform \emph{Human Only}. From Figure \ref{fig:team_performance} (b), we can see that although \emph{AI Confidence} made humans agree with AI more when AI's recommendation was correct, it also made humans agree with AI's wrong recommendation more. This finding is consistent with previous work revealing that showing AI confidence does not necessarily improve team performance \cite{zhang2020effect, rastogi2020deciding}. Compared with \emph{AI Confidence}, the proposed CL exploitation methods achieved marginally to significantly higher team performance (\emph{Direct Display}: $p$=.078; \emph{Adaptive Workflow}: $p$=.027; \emph{Adaptive Recommendation}: $p$=.078).

\textbf{Finding 3: Humans in the proposed three CL exploitation conditions trusted the AI more appropriately when the AI's recommendation was wrong.} As shown in Figure \ref{fig:team_performance} (b) (the red bars), the human-AI agreement in \emph{AI Confidence} was significantly higher than \emph{Adaptive Workflow} ($p$=.017), and \emph{Adaptive Recommendation} ($p$<.001), and it was marginally higher than \emph{Direct Display} ($p$=.07). Note that when AI is wrong, a lower agreement with AI is better. These results suggest humans' \textbf{less over-trust} in AI in our proposed CL exploitation conditions. Through qualitative feedback from participants, we found that our three designs prompted people to rely more on their own thinking when AI's advice was wrong (AI's CL was often lower than human's). Specifically, in \emph{Direct Display}, displaying a higher human CL made them more confident in their own judgment. For example, P26 (49, female, knew basic knowledge of AI) said, ``\emph{Sometimes the AI's opinion differed from mine. When I saw that my (CL) value was higher than the AI's, it confirmed my opinion.}'' While in \emph{Adaptive Workflow}, in most cases, when participants had to make their own judgment first, then did not change their decision later, even if the AI's recommendation was the opposite. For example, P16 (31, male, knew or used AI algorithms) mentioned ``\emph{I had made careful thinking before the AI's suggestion, and I would stick to my own opinion.}'' Similar phenomenon can be found in \emph{Adaptive Recommendation}. P4 (40, female, no knowledge of AI) said, ``\emph{AI didn't tell me any answers, I could only make decisions according to my own thoughts.}''

However, when the AI's recommendation was correct (the red bars in Figure \ref{fig:team_performance} (b)), we did not observe significant differences in human-AI agreement between the three proposed conditions and \emph{AI Confidence} baseline. We infer this might be because the task instances where humans and AI could make correct predictions were highly overlapped. It can be seen from \emph{Human Only} that in the case where AI gave correct advice, even if people did not get any assistance from the AI, their performance also reached ~80\% (agree with AI on 80\% cases), which indicates the complementarity of human and AI in such situations was relatively weak, and the room for improvement was thus limited.




\textbf{Part 2}

\begin{figure*}[h]
\centering
\subfigure[]{
\begin{minipage}[t]{0.46\linewidth}
\centering
\includegraphics[width=\linewidth]{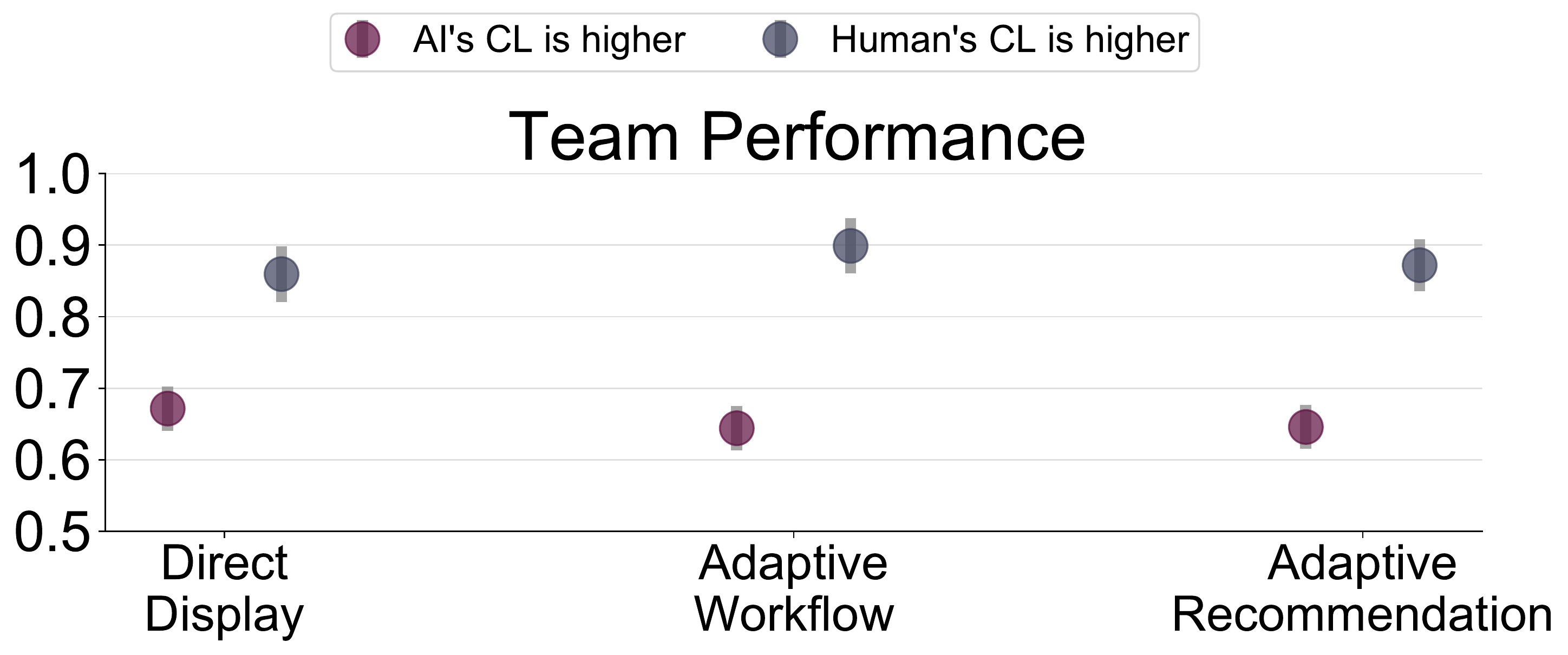}
\end{minipage}%
}%
\subfigure[]{
\begin{minipage}[t]{0.54\linewidth}
\centering
\includegraphics[width=\linewidth]{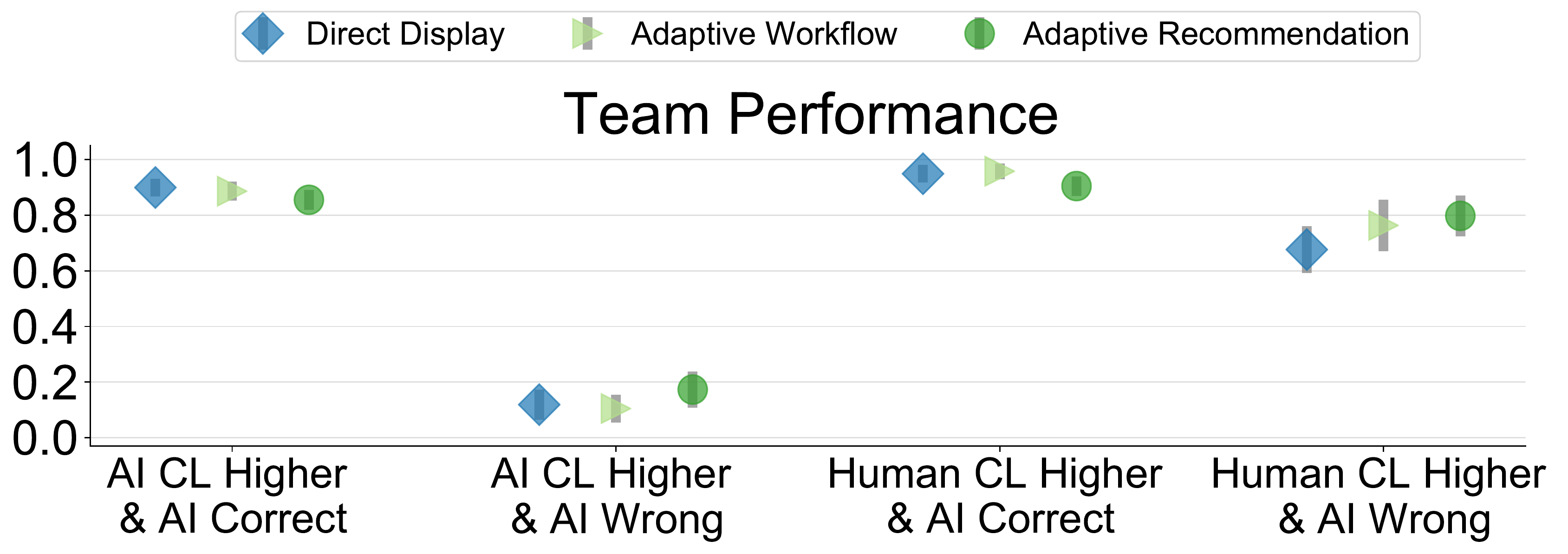}
\end{minipage}%
}%
\centering
	\caption{Team performance (with average accuracy and 95\% confidence interval) in different CL situations. (A) Team performance in different human-AI CL situations under three conditions. (B) We combine the correctness of AI suggestions with human-AI CL to analyze the team performance of the three conditions in different situations in detail.}
	\label{fig:team_performance_CL}   
\end{figure*}

\textbf{Finding 4: Team performance was better when humans' CL was higher.}
In the proposed three CL exploitation conditions, we show participants different information or change the decision-making workflow based on human-AI CL. Therefore, we want to analyze how the team performance differs in different CL situations and different AI recommendation correctness. Overall, as shown in Figure \ref{fig:team_performance_CL} (a), when the human CL was higher, the team performance was significantly better than when the AI's CL was higher ($p$<.001 in all conditions).


Specifically, as shown in Figure \ref{fig:team_performance_CL} (b), (1) when AI's CL was higher \& AI's recommendation was correct, there was no significant difference in team performance between the three conditions. (2) When AI's CL was higher \& AI's recommendation was wrong, there was no significant difference in team performance between the three conditions. (3) When human's CL was higher \& AI's recommendation was correct, compared with \emph{Adaptive Recommendation}, the team performance in \emph{Direct Display} and \emph{Adaptive Workflow} was significantly higher ($p$<.05, $p$<.01 respectively). This might be because in \emph{Adaptive Recommendation} condition, when the human's CL was higher, the AI's suggestions were not shown, thus participants could not get the help of the AI's correct suggestions. (4) When human's CL was higher \& AI's recommendation was wrong, the team performance in \emph{Direct Display} was marginally significantly lower than \emph{Adaptive Recommendation} ($p$<.1) and significantly lower than \emph{Adaptive Recommendation} ($p$<.05).

Furthermore, as expected, we found that when the AI's recommendation was wrong, the team performance when the human's CL was higher was significantly better than when the AI's CL was higher ($p$<.001 in all conditions). 
But to our surprise, when the AI's recommendation was correct, the team performance when the human's CL was higher was significantly better than when the AI's CL was higher, \emph{Direct Display} ($p$<.05); \emph{Adaptive Workflow} ($p$<.01); \emph{Adaptive Recommendation} ($p$=.058, marginally).
The possible reason was that when AI's CL was higher, since human's capability was not as good as AI's, people sometimes listened to their own wrong judgments and thus \emph{under-trust}ed AI.


\textbf{Finding 5: The \emph{Direct Display} and \emph{Adaptive Workflow} conditions worked better when the AI's confidence level ``contradicted'' the correctness of the AI's recommendation.}
There is an insufficiency of only utilizing AI confidence to calibrate humans' trust. Specifically, even if the AI's confidence is higher than a threshold (we used 0.7 \cite{zhang2020effect, wang2021explanations}), AI may still output wrong predictions (denoted as \emph{High \& Wrong region}). Sometimes even if AI's confidence is low, AI can give correct recommendations (denoted as \emph{Low \& Correct region}). We name these two situations as \emph{Conflict region}. We argue that just showing AI's confidence is insufficient for people to recognize these situations.

\begin{figure*}[h]
\centering
\subfigure[]{
\begin{minipage}[t]{0.29\linewidth}
\centering
\includegraphics[width=\linewidth]{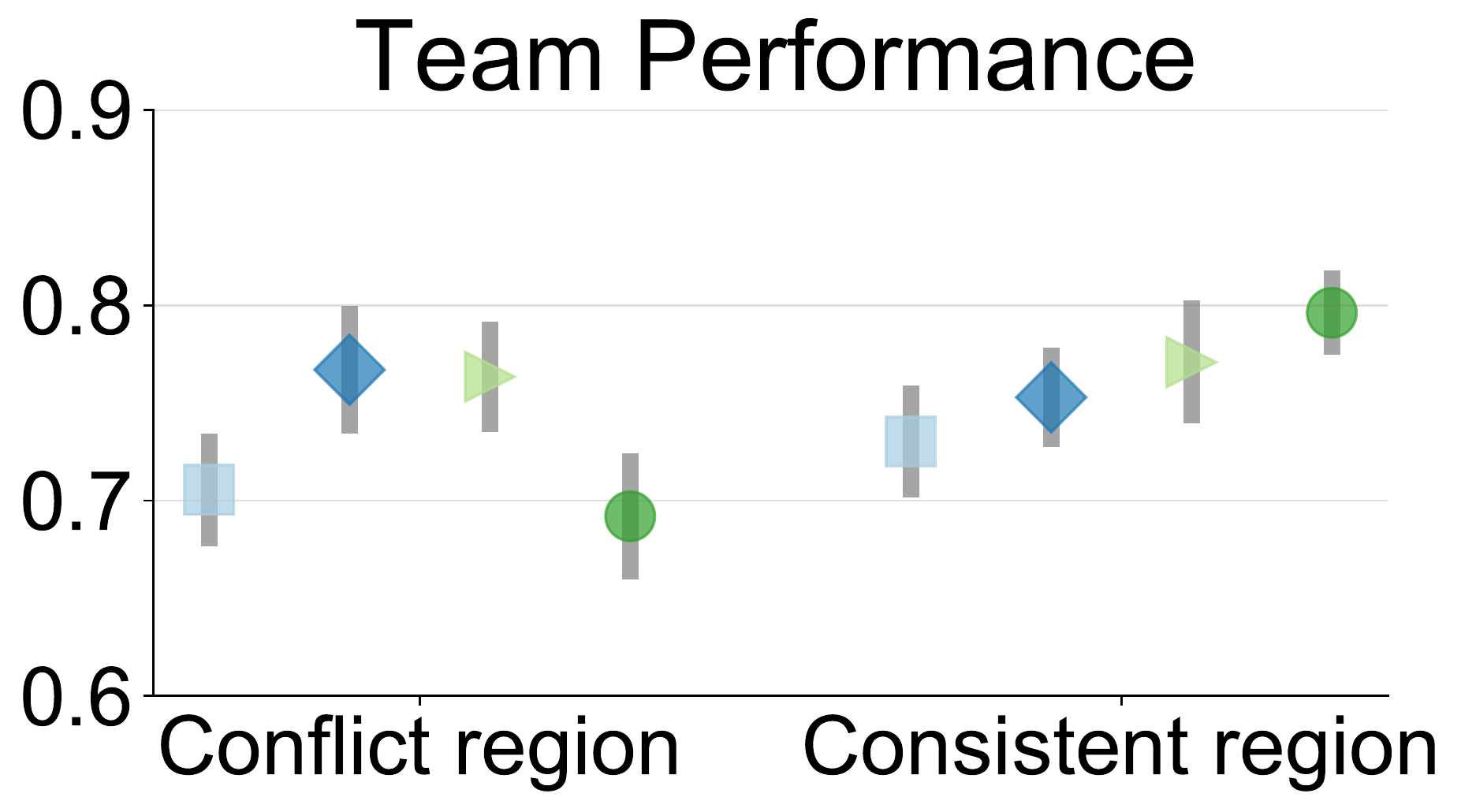}
\end{minipage}%
}%
\subfigure[]{
\begin{minipage}[t]{0.71\linewidth}
\centering
\includegraphics[width=\linewidth]{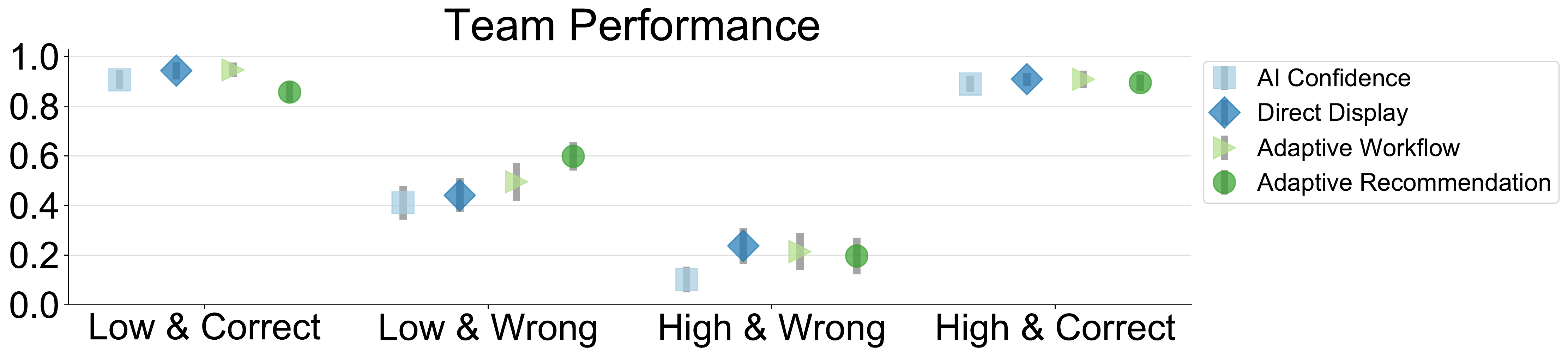}
\end{minipage}%
}%
\centering
	\caption{(A) Team performance (with the mean accuracy and 95\% confidence interval) when the AI's confidence level is in conflict (denoted as \emph{Conflict region}) and consistent (denoted as \emph{Consistent region}) with the correctness of the recommendation given by the AI. (B) Specifically, we divide the \emph{Conflict region} into (1) \emph{Low \& Correct} (AI's confidence is below the threshold but the recommendation is correct) and (2) \emph{High \& Wrong} (AI's confidence is above the threshold but the recommendation is wrong), and we divide the \emph{Consistent region} into (1) \emph{Low \& Wrong} (AI's confidence is below the threshold and the recommendation is wrong) and (2) \emph{High \& Correct} (AI's confidence is above the threshold and the recommendation is correct).}
	\label{fig:four_situation}   
\end{figure*}

In general, as shown in Figure \ref{fig:four_situation} (a), in \emph{Conflict region}, the team performance in \emph{AI Confidence} was significantly lower than \emph{Direct Display} ($p$=.002), \emph{Adaptive Workflow} ($p$=.006). No significant difference was found between \emph{AI Confidence} and \emph{Adaptive Recommendation}.
Specifically, as shown in Figure \ref{fig:four_situation} (b), in the \emph{Low \& Correct region}, there was no significant difference between \emph{Direct Display}, \emph{Adaptive Workflow} and \emph{AI Confidence}. The possible reason may be that the room for improvement is limited (already exceeds 90\%). We noted that \emph{Adaptive Recommendation} was significantly lower than \emph{AI Confidence} ($p$=.032) probably because, in \emph{Adaptive Recommendation} condition, people developed a mode of independent thinking without relying on AI advice because they often could not see AI advice, which might lead to \emph{under-trust}. In the \emph{High \& Wrong region}, team performance in \emph{AI Confidence} was significantly lower than \emph{Direct Display} ($p$=.005), \emph{Adaptive Workflow} ($p$=.022) and marginally lower than \emph{Adaptive Recommendation} ($p$=.056).


\textbf{Finding 6: The \emph{Adaptive Workflow} and \emph{Adaptive Recommendation} conditions worked better when the AI's confidence level was ``consistent'' with the correctness of the AI's recommendation.} Another category of task instance is called \emph{Consistent region}, which includes (1) \emph{Low \& Wrong} (when the AI's confidence is below the threshold and the recommendation given is wrong), and (2) \emph{High \& Correct} (when the AI's confidence is above the threshold and the recommendation given is correct).

In general, as shown in Figure \ref{fig:four_situation} (a), the team performance in \emph{AI Confidence} was marginally significantly lower than \emph{Adaptive Workflow} ($p$=.074) and significantly lower than \emph{Adaptive Recommendation} ($p$<.001). But no significant difference can be found between \emph{AI Confidence} and \emph{Direct Display}. We then dig deeper into the two sub-regions. In the \emph{Low \& Wrong region}, the team performance in \emph{AI Confidence} was marginally significantly lower than \emph{Adaptive Workflow} ($p$=.088) and significantly lower than \emph{Adaptive Recommendation} ($p$<.001). But no significant differences were observed between \emph{AI Confidence} and \emph{Direct Display}. In the \emph{High \& Correct region}, there was no significant difference between the \emph{AI Confidence} baseline and any of the proposed three CL exploitation conditions.


In addition to the above results, we also found that (1) the proposed three conditions effectively conveyed the estimated CL information to humans, and (2) participants performed better on \emph{Consistent CL} task instances (where their perceived human-AI CL was consistent with the system's communicated human-AI CL). Detailed results can be found in the supplementary material.

\textbf{Summary.} Overall, our proposed three CL exploitation methods promoted humans' appropriate trust in AI (especially reducing humans' over-trust and without causing under-trust) and thus led to better team performance. Also, the proposed three CL exploitation strategies could effectively communicate the system's estimated human-AI CL to humans. In addition, our methods outperformed the \emph{AI Confidence} baseline when the AI's confidence contradicted its correctness. We also notice some pitfalls in our designs and will discuss them in the later section.


\subsection{Effects of CL Exploitation Strategies on Human Perceptions and Experiences.}
\begin{figure*}[t]
	\centering 
	\includegraphics[width=1\linewidth]{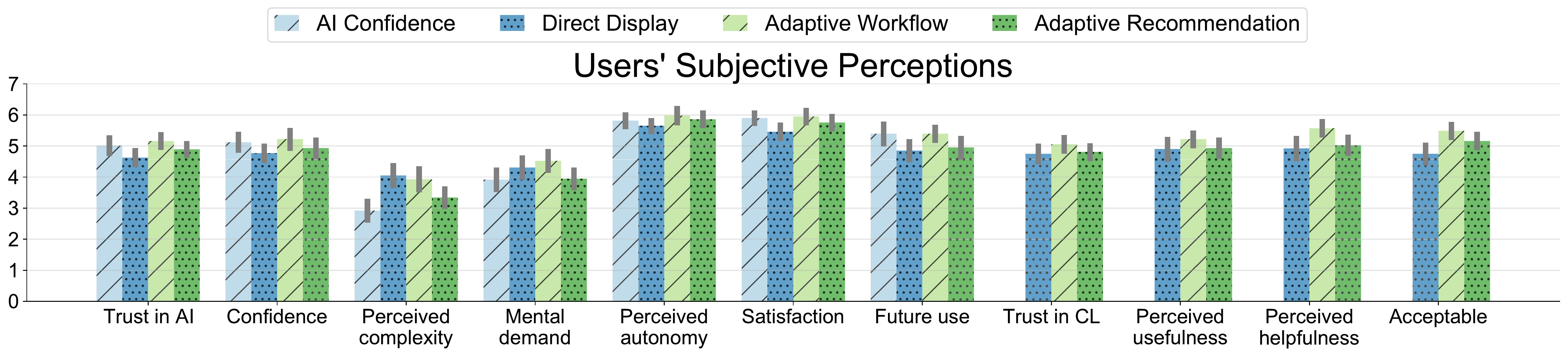}
	\caption{Participants' subjective ratings in the exit survey (with the mean values and 95\% confidence interval).}
	\label{fig:subjective}
\end{figure*}

To answer RQ3, we analyze participants' subjective perceptions in different conditions in the exit survey, with a 7-point Likert scale (1: Strongly disagree, 7: Strongly agree). Figure \ref{fig:subjective} shows the results.


\textbf{Perceived complexity of the system.} Overall, participants' perceived system complexity is relatively low to neutral in the four conditions. Kruskal-Wallis test reveals significant differences among different conditions ($\chi^2$=19.223, $p$<.001). Post-hoc analysis shows that compared with \emph{AI Confidence}, participants found the system significantly more complex in \emph{Direct Display} ($p$<.001) and \emph{Adaptive Workflow} ($p$<.01), perhaps because the two conditions display more information and require more complex workflow.

\textbf{Mental demand.} Overall, participants were neutral about whether the decision-making process was mentally demanding in the four conditions. There are no statistically significant differences among different conditions. However, we observe a trend that \emph{Adaptive Workflow} leads to high mental demand for participants.

\renewcommand{\arraystretch}{1.2}
\begin{table*}[tp]  
  \centering  
  \fontsize{8}{8}\selectfont  

\caption{Participants' qualitative feedback in the open-ended questions. (Note that we excluded answers that just gave positive feedback but not specific, such as ``helpful'', ``like it'', so the sum of the participants may not reach 100\%.)}\label{table:openend} 
\begin{tabular}{p{2.5cm}p{10cm}p{1.5cm}}
\toprule
Themes&Definitions and Examples&\#Participants\\
\midrule
\multicolumn{3}{l}{\textbf{How did participants perceive the estimated human-AI CL (in \emph{Direct Display})?}}\\
\emph{Doubt the CL}&Some participants did not believe their abilities could be easily and accurately estimated.&14 (23\%)\\
&``\emph{It was just a guess. The AI did not actually know about me so it did not seem reliable.}'' (P41)&\\
\cline{2-3}
\emph{Feeling of confirmation}&If the human and AI had the same views and the human's CL was high, it made participants more confident.&8 (13.5\%)\\
&``\emph{It made me feel very confident on those that we all agreed on}'' (P3)&\\
\cline{2-3}
\emph{Realize AI's flaws}&When people saw AI's lower CL, they realized that AI was not always trustworthy.&6 (10\%)\\
&``\emph{It made me recognize the AI could also be flawed on this task. I was better than the AI!}''(P5)&\\
\cline{2-3}
\emph{Decrease confidence}&Sometimes people became less confident when the displayed CL of them was low.&5 (8\%)\\
&``\emph{I began to question my capability somewhat when I viewed my estimated capability that was displayed low in some cases.}'' (P14)&\\
\midrule

\multicolumn{3}{l}{\textbf{How did participants use the displayed human-AI CL to make a decision (in \emph{Direct Display})?}}\\
\emph{Rely on the higher one}&They first looked at the two CL values. If AI's was higher, they followed AI's recommendation. Otherwise, they would make their own decisions.&19 (32\%)\\
&``\emph{The two charts (displaying human-AI CL) showed me when I should go with my own gut instincts and when I should rely on the AI instead.}''(P12)&\\
\cline{2-3}
\emph{Just ignore the CL}&Some participants only believe in themselves and completely ignore the CL. &15 (25\%)\\
&``\emph{I trusted my capability more, and did not put much stock in the displayed values}''(P23)&\\

\cline{2-3}
\emph{Reflect upon it}&Some participants reflected their decisions after seeing the estimated CL.&10 (17\%)\\
&``\emph{It helped me reflect on my decision once I saw my score was not as high as I thought.}''(P12)&\\

\cline{2-3}
\emph{Refer to it in inconsistent cases}&Some participants first made their own decisions. If AI agreed with them, they would ignore the CL. Otherwise, they mainly listened to the party with the higher CL.&9 (15\%)\\
&``\emph{I directly made the decision if I saw I and AI were the same. If the AI disagreed with me, I compared our abilities and chose the higher one to follow.}''(P49)&\\

\cline{2-3}
\emph{Refer to it on uncertain cases}&In cases where people were not confident, they would refer to the CL. And when they are confident, they would ignore it.&6 (10\%)\\
&``\emph{It did help me make some decisions where I was unsure. However, I would not consider it when I felt I was totally correct.}''(P39)&\\

\midrule
\multicolumn{3}{l}{\textbf{How were participants' decision processes influenced by the adaptive workflow?}}\\
\emph{Devote more cognitive resources}&When people were asked to make decisions first, they devoted more cognitive resources to the task itself, avoiding being influenced by AI's judgment.&30 (53\%)\\
&``\emph{I paid closer attention to what I thought was correct if the AI didn't make a recommendation first. If it did, then I more or less yielded to the AI's judgment.}''(P42)&\\
\cline{2-3}
\emph{Little influence}&Some participants regarded AI as a double-check or second opinion.&26 (46\%)\\
&``\emph{It did not affect me much. I always treated the AI as a second opinion, no matter whether the AI allowed me to decide first or not.}''(P54)&\\

\midrule
\multicolumn{3}{l}{\textbf{How were participants' decision processes influenced by the adaptive AI recommendation?}}\\
\emph{Independent thinking}&Not showing AI's concrete recommendation required people to think independently.&35 (62\%)\\
&``\emph{When I looked at the AI's recommendation, it was harder to trust myself and instead I found myself defaulting to the AI.}''(P33)&\\
\cline{2-3}
\emph{Little influence}&It did not affect their decision because they often relied on themselves.&19 (33\%)\\
&``\emph{It didn't affect my decision-making. I was still thinking about each information card that is presented.}''(P44)&\\

\bottomrule
\end{tabular}
\end{table*}




\textbf{Perceived helpfulness to decide when to trust the AI.} Overall, participants thought the estimated human-AI CL was helpful to make a trust choice. Kruskal-Wallis test reveals significant difference among different conditions ($\chi^2$=7.039, $p$<.05). Post-hoc analysis shows that participants in \emph{Adaptive Workflow} found the estimated human-AI CL marginally more helpful than \emph{Direct Display} ($p$=.056) and \emph{Adaptive Recommendation} ($p$=.073).

\textbf{Acceptance of estimation of their CL.} Overall, participants thought the estimation of their CL is acceptable. Kruskal-Wallis test reveals significant difference among different conditions ($\chi^2$=9.162, $p$<.01). Post-hoc analysis shows that compared with \emph{Direct Display} condition, participants in \emph{Adaptive Workflow} thought the estimation of their CL significantly more acceptable ($p$<.01).

However, in terms of \textbf{Trust in AI}, \textbf{Confidence in the decision-making process}, \textbf{Perceived autonomy}, \textbf{Satisfaction}, \textbf{Future use}, \textbf{Trust in the estimation of human-AI CL}, and \textbf{Perceived usefulness of estimation of human-AI CL}, there is no significant difference among different conditions.

In summary, except for the perceived system complexity, our proposed three conditions did not cause significantly different perceptions of participants on other aspects. Within our proposed three CL exploitation methods, \emph{Adaptive Workflow} seems to earn people's higher perceptions regarding its helpfulness and acceptance. This also echoes the result that \emph{Adaptive Workflow} achieved the highest team performance among all conditions. However, we also note that the benefits come with increased complexity and mental demand. Therefore, future work is suggested to explore a trade-off between effectiveness and user experience with human-involved empirical studies.

\subsection{Qualitative Analysis on How Participants Perceived, Used, and Were Affected by the Human-AI CL.}
To better understand the effects of different CL exploitation conditions, in the exit survey, we leave open-ended questions asking how participants perceived and utilized the human-AI CL (in \emph{Direct Display}) and were influenced by the adaptive decision-making process (in \emph{Adaptive Workflow} and \emph{Adaptive Recommendation}). Following an inductive thematic analysis process \cite{hsieh2005three} (Sec. \ref{analysismethod}), two authors created a codebook (Table \ref{table:openend}). We highlight the following findings which explain the quantitative results afore-mentioned.

\textbf{Some participants doubted the displayed CL information and ignored it in the decision-making process.}
In \emph{Direct Display}, 23\% of participants doubted the displayed CL, and 25\% ignored the CL in the decision-making process.
This also echoes the results that even though the AI's CL was higher, the human-AI agreement still did not reach 100\% (Figure \ref{fig:agreement}), and it might explain why when AI' CL was higher and AI's recommendation was correct, the team performance still did not reach 100\% (Figure \ref{fig:team_performance_CL} (b)).

\textbf{Most participants referred to or were influenced by the displayed CL or CL-based adaptation.} 
In \emph{Direct Display} 74\% of the participants referred to CL, and in \emph{Adaptive Workflow} and \emph{Adaptive Recommendation}, 53\% and 62\% of participants were affected by the adaptive process respectively. This is consistent with the results that the three proposed methods could effectively affect people's agreement with AI (Figure \ref{fig:agreement}), and promote humans' appropriate trust (Figure \ref{fig:team_performance}).

\textbf{Participants in \emph{Adaptive Workflow} and \emph{Adaptive Recommendation} were forced to think independently.}
In \emph{Adaptive Workflow} and \emph{Adaptive Recommendation}, most participants were influenced by the adaptive process to think independently when human CL is higher. This supports the reason why our methods helped reduce over-trust when AI was wrong (Figure \ref{fig:team_performance} (b)). In particular, 62\% of participants in \emph{Adaptive Recommendation} would think independently when they could not see an AI recommendation, and 33\% said they always thought on their own. This reflects that participants formed a pattern of not relying on the AI, so the human-AI agreement is lowest when the AI's CL is higher (Figure \ref{fig:agreement}). This also explains why team performance is the lowest in \emph{Adaptive Recommendation} in the \emph{Low \& Correct} region (Figure \ref{fig:four_situation}). For other findings please refer to our codebook (Table \ref{table:openend}).

%% file: 06-Discussion.tex
\section{Discussion}

Through investigating the three research questions, our study shows the promise of modeling and communicating humans' CL for promoting appropriate human trust in AI-assisted decision-making. Based on our main findings, we discuss several key issues for improving decision-making with human-AI teams and the limitations of our work.


\subsection{Human Perceptions of Self-confidence and Understanding of AI's confidence}

\textbf{Maintaining proper self-confidence is critical for humans to establish appropriate trust in AI.} Evidence shows that people's confidence in themselves significantly affects whether they will take AI's advice \cite{vodrahalli2022humans, chong2022human}. However, individuals' confidence in their own capabilities may mismatch with their actual capabilities \cite{moore2020perfectly, miller2015meta, weber2004confidence, meyer2013physicians, kahneman2011thinking} for both experts and lay people, leading to overconfidence (or underconfidence) \cite{turnercalibrating, meyer2013physicians, weber2004confidence, miller2015meta}.
From our results, in \emph{AI Confidence} condition, when AI offered correct recommendations (also with high confidence), some participants, however, still followed their wrong judgments. This is because it is difficult for humans to maintain a ``\emph{calibrated}'' self-confidence \cite{moore2020perfectly}, thus overlooking AI's suggestions. We believe that if people could accurately perceive their abilities (e.g., correctness likelihood) and calibrate self-confidence accordingly, the collaboration between humans and AI will be more successful. 
The human CL modeling and communication method proposed in this paper is an initial step toward this goal.
We hope our work can inspire researchers to explore effective ways to adjust humans' confidence in AI and in themselves with human-centered computational modeling and interface design.

\textbf{Humans' understanding of probability affects the effectiveness of trust calibration.} Both the AI's confidence and the human's CL are numerical probabilities. However, previous works suggest that it is difficult for humans to act on numbers (e.g., confidence, accuracy) to coordinate the efforts of AI, especially with limited cognitive resources in some time-critical scenarios \cite{buccinca2020proxy, berwick1981doctors, lai2019human, slovic2006risk}. Furthermore, people, especially who are not good at applying mathematical thinking, lack the ability to easily interpret what a probability value actually means \cite{cosmides1996humans, peters2006numeracy, reyna2008numeracy}. This is possibly one reason why displaying AI's confidence score to humans is insufficient for calibrating their trust, which is revealed by both our work and existing studies \cite{zhang2020effect, rastogi2020deciding}.
Thus, it can be challenging to rely solely on people to make rational reliance choices and coordination (e.g., \emph{who makes a prediction first}). Our work proposes leaving the computational probability estimation/comparison task to the system and calibrating human trust by automatically adapting the decision-making process/interface. This can counter possible human cognitive biases \cite{bertrand2022cognitive, wang2019designing} and avoid making people directly deal with probabilities. Future work could explore two other directions. One is to design more effective algorithm-in-the-loop task coordination methods (e.g., \emph{learning to defer} \cite{madras2018predict}) while retaining a proper level of human autonomy. The other is to design interfaces to improve people's comprehension of probabilities, such as adding a simple tutorial about probability and frequency \cite{moore2017confidence}, presenting probabilities in more understandable manners \cite{lai2019human}, etc.



\subsection{Achievement of Complementary Performance beyond Trust Calibration}


\textbf{Exploiting knowledge complementarity is beneficial for team performance.} Our study found that although the team performance in the proposed three conditions exceeded \emph{AI Only} and \emph{Human Only}, the improvement was not ``remarkable'' (about 3-4\%). One of the key reasons is that the \textbf{complementary region/zone} between humans' knowledge space and that of AI's is relatively small.
It is reflected by the performance analysis in \emph{Human Only} that there are only a few instances that only one member of the human-AI team can handle correctly, making it hard to achieve substantial complementary performance just by calibrating human trust.
In comparison, in Bansal et al.'s work \cite{bansal2021does} where complementary performance is achieved, humans' independent accuracy is even higher when AI \emph{cannot} gives a correct recommendation than when AI \emph{can}. Therefore, echoing \cite{bansal2021does, zhang2020effect}, to ensure complementary performance, besides calibrating people's trust in AI, it is necessary to harness the complementarity of human and AI intelligence to achieve optimal outcome \cite{bansal2019updates, wilder2020learning}, perhaps by training an AI that can complement humans' knowledge and error regions. 

\textbf{The modeling of human capability can empower more elaborate designs.}
In addition to approximating people's CL, our proposed modeling method is able to estimate people's predictions. We think this information can be valuable because it can help us project in advance whether humans will make consistent judgments with AI. We can combine this information with human-AI CL to enable more sophisticated strategies for assisting humans in making better decisions. For example, when the judgments of humans and AI are predicted to be consistent and neither of their CL is high, both of them are likely to make a wrong prediction. In such a case, AI can focus on encouraging people to think analytically rather than affirming people's decisions. For example, as suggested in \cite{bansal2021does}, AI sometimes can play the role of devil's advocate and question humans' judgment. Future work could explore more advanced approaches to leveraging humans' decision-making models and human-AI capabilities proposed in this paper.


\subsection{Design of Appropriate CL Communication Methods}

\textbf{Appropriately communicating the CL information is as important as correctly modeling it.} Because humans are the ultimate decision-makers, how people receive, perceive, and use this information in their decision process is essential to the outcome. Although the three CL exploitation mechanisms proposed in this paper improved people's appropriate trust in AI, we found that there were still some participants who held onto their misjudgments.
From the open-ended feedback, we found that some participants did not think that their ability could be easily and reliably estimated by the system, which undoubtedly hindered the potential of our method. Although we had told them the necessary information,
the underlying process was still a kind of \emph{black box} to some participants. Thus, we suggest providing a more detailed and easy-to-understand guide to introduce and explain the rationale behind the CL modeling to humans, increasing their understanding and acceptance.


\textbf{Other potential effective CL communication designs.} 
Besides the proposed adaptive design, other types of information may also be adapted to facilitate the calibration of human trust. Previous works show that the availability of AI's explanations, regardless of their correctness, is likely to increase people's trust in AI \cite{wang2021explanations, bansal2021does, poursabzi2021manipulating, lai2019human}.
Hence, we may design an \emph{Adaptive Explanation} strategy to provide AI's explanation only when AI's CL is higher than humans'. 
In addition, some studies found that the framing of confidence may affect people's perception of risk \cite{christopoulos2009neural}. We thus can apply a positive tone to describe the AI's CL when it is high, e.g., ``AI has a 75\% chance to make a correct prediction'', and use an uncertain tone otherwise, e.g., ``AI has a 25\% chance to make a wrong prediction'' (although equivalent to the former). 
Besides, recent work highlights the dual-process of cognition when people process information in decision-making \cite{buccinca2020proxy, kahneman2011thinking, cacioppo1984elaboration, wason1974dual}, where System 1 processes stimuli in a fast and automatic manner which could lead to cognitive biases if applied inappropriately \cite{kahneman2011thinking}, whereas System 2 engages in deliberative and analytical thinking.
One general way to leverage this theory is when people need to rely more on their own judgment such as when their CL exceeds the AI's, the interface should stimulate people's System 2 thinking. Through these theoretical lenses, we can design more effective usage of CL information.

\subsection{Pitfalls of Current CL Modeling and Exploitation Methods}

\textbf{Potential side effects of the interface.} 
Despite the effectiveness of our proposed designs in promoting humans' appropriate trust, we still suggest designers be cautious of their potential pitfalls. For example, in \emph{Adaptive Recommendation}, we found that the participants seemed to form a pattern of \textbf{skepticism} of AI because they often could not see AI suggestions, which might hinder their utilization of AI's assistance when AI's correct advice is shown. In addition, \emph{Adaptive Workflow} may lead to humans' confirmation bias \cite{nickerson1998confirmation}. For example, after people made an initial judgment and then found AI's ``confirmation'', they would be very sure that this was the correct answer. But in fact, sometimes people and AI make wrong judgments simultaneously. Therefore, we recommend that, in addition to grounding a design in existing cognitive theories, it is necessary to verify the potential impact and adverse effects of the design empirically.


\textbf{The drawback of human-AI CL and its ethical and accountability issues.}
There are two issues surrounding human-AI CL. First, even if the AI's confidence is calibrated and the human's CL is accurately modeled, inconsistent cases still exist: In the human-AI team, for a specific task instance, the member with higher CL makes a wrong prediction while the member with lower CL makes a correct prediction. It may lead to humans' inappropriate trust in AI. Since CL is just a \emph{probability} of being correct, such \emph{inconsistency of CL and correctness} is inevitable. Second, our estimation of human CL can be imperfect, and an AI model's confidence can sometimes be poorly-calibrated. So, using human-AI CL inappropriately may lead to negative results. For example, if we mistakenly estimate a human's CL to be lower, our method may lead the human to accept the wrong advice from AI when she/he could have made a correct decision independently. It can induce severe consequences and even become dangerous in high-risk scenarios. Therefore, for human-AI CL to play a positive role, confirming the reliability of human-AI correctness likelihood is essential before deployment. Besides, it may be beneficial to communicate the uncertainty behind the CL wherever appropriate to warn human decision-makers of the risk of such information.
Another possible way to mitigate the negative impact of the estimated CL is to avoid conveying a sense of confirmed, precise information, such as using specific percentages or judgmental words \cite{reyes1980judgmental}. Instead, researchers could communicate this information implicitly, embedding it in the decision-making process through designs similar to our proposed adaptive methods. 





\subsection{On the Generalizability of Our Method and Results}
Proper caution should be used when generalizing our method and results to different task domains and subject populations. First, we choose a rule-based approach to help users understand and modify the auto-generated decision model. However, this approach may not be suitable for more complex decision tasks such as those involving text or image data. Thus, we need to design proper knowledge representation and modeling algorithms based on the specific characteristics of the task and data. For example, in a textual sentiment analysis task, users can specify keywords or example sentences to represent their decision model \cite{lai2022human}.
Second, our study was conducted on non-expert users in low-stake decision-making tasks. While it is a suitable testbed for exploring humans' trust appropriateness in AI-assisted decision-making \cite{zhang2020effect, ghai2021explainable}, we caution readers to generalize our results to other populations or other tasks. For example, it is unclear whether our results will still hold when our designed interfaces are adopted in high-stake tasks (where the decision-maker might have different cognitive routes \cite{suresh2021beyond}). And whether domain experts' capabilities can be well modeled by our method is also unknown. Nevertheless, we believe our proposed framework to calibrate humans' trust based on both sides' capabilities can be generalized to different AI-assisted decision-making scenarios where collaboration is needed. Future work can adapt our human CL modeling and communication method to other decision-making tasks with different stakeholders \cite{liao2021human}.

\subsection{Limitations and Future Work}
\label{limitation}
There are several limitations in our proposed methods and experimental setting. First, we used decision rules to approximate humans' decision-making models. However, rules only provide a general model and cannot cover all edge cases. Future solutions can consider integrating the ``\emph{behavioral testing}'' method \cite{beizer1995black} where the system can use test cases to ``check'' users' ability, just like testing a software or NLP model \cite{ribeiro2020beyond}. Second, we did not update humans' decision-making models in the last 20 tasks because we focused on studying the impact of our method on humans in the scope of this paper. We assume that in the absence of correctness feedback (e.g., no access to ground truth), the user's decision model is relatively fixed in the short term, which is reasonable for our experiments. However, in real-world decision-making, users' decision-making models can change as users interact with AI services and encounter more task instances \cite{ma2019smarteye}, so a static model is not enough. In the future, we plan to explore how to maintain a real-time updated user decision model in long-term AI-assisted decision-making. Third, we measured human trust in AI by human-AI agreement. Although it is widely used \cite{zhang2020effect, wang2021explanations,bansal2021does}, an obvious shortcoming is that, when people's final judgment is consistent with the AI's, we cannot distinguish whether it is because they listened to the AI's advice or because their own decisions are consistent with the AI's. Future studies may explore more suitable measurements.

\section{Conclusion}
Humans' appropriate trust in AI is a fundamental challenge in AI-assisted decision-making, and our work makes a contribution toward calibrating humans' trust based on the capabilities of both humans and AI. Our investigation consists of two consecutive phases. In the first phase, we explore how to model humans' capability (correctness likelihood) on a given task instance. We propose a human decision-making model approximation method with an interactive decision rule modification interface. In the second phase, we explore how to leverage human-AI capabilities to promote appropriate trust in AI-assisted decision-making. Based on theories of people's cognitive processes, we propose three CL exploitation methods and investigate their effects on humans' trust appropriateness, task performance, and user experience. Our results highlight the effectiveness of the proposed human CL modeling and exploitation method in promoting more appropriate human trust in AI compared with the traditional AI confidence-based method. With the derived practical implications based on our main findings, we hope this work to be a step towards promoting appropriate human-AI decision-making by considering the mutual capability information of both sides.